\documentclass[prb,amsmath,amssymb,twocolumn,lengthcheck,showpacs,floatfix,groupedaddress,superscriptaddress]{revtex4}
\usepackage{graphicx}
\usepackage[english]{babel}
\usepackage{times}
\usepackage{dcolumn}
\usepackage[usenames,dvipsnames]{color}
\usepackage{units}

\usepackage{color}

\begin{document}

\title{Pad\'e approximants for improved finite-temperature spectral
functions in the numerical renormalization group}

\author{\v{Z}iga Osolin}

\affiliation{Jo\v{z}ef Stefan Institute, Jamova 39, SI-1000 Ljubljana,
Slovenia}

\author{Rok \v{Z}itko}

\affiliation{Jo\v{z}ef Stefan Institute, Jamova 39, SI-1000 Ljubljana,
Slovenia}
\affiliation{Faculty  of Mathematics and Physics, University of Ljubljana,
Jadranska 19, SI-1000 Ljubljana, Slovenia}

\date{\today}

\begin{abstract}
We introduce an improved approach for obtaining smooth
finite-temperature spectral functions of quantum impurity models using
the numerical renormalization group (NRG) technique. It is based on
calculating first the Green's function on the imaginary-frequency
axis, followed by an analytic continuation to the real-frequency axis
using Pad\'e approximants. The arbitrariness in choosing a suitable
kernel in the conventional broadening approach is thereby removed and,
furthermore, we find that the Pad\'e method is able to resolve fine
details in spectral functions with less artifacts on the scale of
$\omega \sim T$. We discuss the convergence properties with respect to
the NRG calculation parameters (discretization, truncation cutoff) and
the number of Matsubara points taken into account in the analytic
continuation. We test the technique on the the single-impurity
Anderson model and the Hubbard model (within the dynamical mean-field
theory). For the Anderson impurity model, we discuss the shape of the
Kondo resonance and its temperature dependence. For the Hubbard model,
we discuss the inner structure of the Hubbard bands in metallic and
insulating solutions at half-filling, as well as in the doped Mott
insulator. Based on these test cases we conclude that the Pad\'e
approximant approach provides more reliable results for spectral
functions at low-frequency scales of $\omega \lesssim T$ and that it
is capable of resolving sharp spectral features also at high
frequencies. It outperforms broadening in most respects.
\end{abstract}


\maketitle

\newcommand{\vc}[1]{{\mathbf{#1}}}
\newcommand{\braket}[2]{\langle#1|#2\rangle}
\newcommand{\expv}[1]{\langle #1 \rangle}
\newcommand{\corr}[1]{\langle\langle #1 \rangle\rangle}
\newcommand{\ket}[1]{| #1 \rangle}
\newcommand{\Tr}{\mathrm{Tr}}
\newcommand{\kor}[1]{\langle\langle #1 \rangle\rangle}
\newcommand{\degg}{^\circ}
\renewcommand{\Im}{\mathrm{Im}}
\renewcommand{\Re}{\mathrm{Re}}
\newcommand{\GG}{{\mathcal{G}}}
\newcommand{\atanh}{\mathrm{atanh}}
\newcommand{\sgn}{\mathrm{sgn}}
\renewcommand{\bm}[1]{\mathbf{#1}}

\section{Introduction} 

Nonmagnetic host metal doped with magnetic impurities exhibits
low-temperature anomalies in its thermodynamic and transport
properties known as the Kondo effect
\cite{hewson,kondo1964,kouwenhoven2001}. The resistance of such
materials has a minimum at finite temperature (the Kondo temperature,
$T_K$) \cite{hewson}. An unyielding mystery for a number of years,
this ``Kondo problem'' of the resistance minimum was eventually
theoretically explained by J. Kondo by showing that the perturbation
theory in higher order applied to the $s$-$d$ exchange Hamiltonian
includes logarithmically divergent terms \cite{kondo1964}. 
The effective exchange interaction strength $J$ is renormalized to
larger values as the temperature is reduced: magnetic impurities are
weakly interacting at high temperatures, but become strong electron
scatterers on the scale of $T_K$
\cite{anderson1969exact1,anderson1970exact2,anderson1970,haldane1978}.
The system behaves as a renormalized local Fermi liquid (FL) at low
temperatures \cite{nozieres1974,hewson1993}. Accurate results
for the temperature dependence of the anomalies in the thermodynamic
properties were obtained by K. Wilson using a novel technique,
the numerical renormalization group (NRG)
\cite{wilson1975,krishna1975,krishna1980a,krishna1980b,bulla2008}.
It consists in reducing the Hamiltonian for the point-like impurity
to its one-dimensional form, discretizing the continuum of
conduction-band electrons in a logarithmic way around the Fermi level,
rewriting the Hamiltonian in the form of a tight-binding chain with
exponentially decreasing hopping matrix elements, and 
numerically diagonalizing the resulting Hamiltonian in an iterative
way. The NRG results for the thermodynamics of the Kondo model were
later confirmed by the Bethe Ansatz approach, which provides an
analytic solution to the problem
\cite{andrei1983,tsvelick1983,tsvelick1983td}.

While thermodynamic properties are simple to compute using the NRG
\cite{wilson1975,krishna1980a,oliveira1981,oliveira1994}, dynamical
properties (spectral functions and various dynamical susceptibility
functions) and transport properties (conductance, thermopower) require
significantly more effort
\cite{oliveira1981phaseshift,frota1986,sakai1989,yoshida1990,costi1991,sakai1992siam,costi1993,costi1994,andergassen2011,rejec2012}.
There are two main difficulties. The first concerns the actual
calculation of the raw spectral functions. A number of algorithmic
improvements over the years have increased the reliability of the
method
\cite{bulla1998,hofstetter2000,campo2005,anders2005,anders2006,peters2006,resolution},
also at finite temperatures \cite{weichselbaum2007}. Since the NRG
calculations are performed for tight-binding chains of finite length,
the raw spectral functions are represented as sets of
weighted delta peaks, and the second difficulty consists in
obtaining a smooth continuous representation of the spectra. At zero
temperature, one usually performs spectral broadening using a
log-Gaussian kernel which is well adapted to the logarithmic
discretization grid \cite{bulla2008}. With high-quality raw results
and using the $z$-averaging trick, it is possible to obtain highly
accurate final results with few overbroadening effects
\cite{resolution,zitko2011,freyn2009}. At finite temperatures, a simple
log-Gaussian broadening kernel is not appropriate for $\omega \lesssim
T$ (in units of $\hbar=k_B=1$). Instead, on low energy scales one
should switch to a 
Gaussian or Lorentzian kernel \cite{bulla2001}. Even when smooth
cross-over functions are used to glue the broadening kernel for
$\omega \gtrsim T$ with that for $\omega \lesssim T$
\cite{weichselbaum2007}, the resulting spectral functions exhibit
artifacts in the cross-over frequency region which cannot be fully
eliminated. It is thus desirable to devise new approaches for
obtaining more accurate continuous representations of spectral
functions.

Impurity problems have applications ranging from magnetically doped
materials and heavy-fermion compounds \cite{coleman2002,allen2005},
electron transport in nanostructures (quantum dots
\cite{glazman1988,goldhabergordon1998b,goldhabergordon1998a,cronenwett1998,wiel2000,pustilnik2004,potok2007,andergassen2010},
carbon nanotubes \cite{nygard2000}, molecules and single-atom
transistors \cite{park2002,liang2002,parks2010,scott2010}, molecular
magnets \cite{romeike2006,romeike2006b}), to spectra of single atoms
probed using a scanning tunneling microscope (STM)
\cite{madhavan1998,neel2007,pruser2011} and dissipative two-state
systems \cite{leggett1987}, but their importance has also vastly
increased in recent decades because they play a central role in the
dynamical mean-field theory (DMFT) for strongly correlated electron
materials
\cite{metzner1989,mullerhartmann1989,kotliar2005,maier2005,kotliar2006,Georges:2011hr}.
In the DMFT, a lattice problem of correlated electrons (such as
Hubbard model, periodical Anderson model, or Kondo lattice model) is
mapped onto an effective single-impurity problem subject to
self-consistency conditions. The approach is based on the observation
that in the limit of infinite dimensions or infinite lattice
connectivity, the self-energy function $\Sigma(\vc{k},\omega)$ becomes
purely local \cite{mullerhartmann1989} and depends only on frequency
$\omega$, not on momentum $\vc{k}$. In this limit the method is exact,
but it is also a good approximation for three-dimensional and some
two-dimensional systems where spatial correlations are less important.
Many techniques have been used in the past to solve the effective
impurity models, which is the most computationally demanding part of
DMFT calculations. Presently, the most popular methods are exact
diagonalization, quantum Monte Carlo, especially the
continuous-time algorithm (CT-QMC)
\cite{werner2006,haule2007,gull2011}, and NRG. The QMC is a
stochastic simulation which is numerically exact, but time consuming.
The results for spectral functions are accumulated in bins defined on
the Matsubara imaginary-frequency axis and to obtain the spectra on
the real-frequency axis one has to perform an analytic continuation
\cite{jarrell1996}, for instance using the maximum entropy method
\cite{jarrell1996} or with Pad\'e approximants
\cite{Vidberg:1977vo,PhysRevB.61.5147}. This procedure is not without
peril and high quality numerical results are required to obtain
reliable spectra. 
The NRG, on the other hand, can provide the results directly on the
real-frequency axis and is thus well adapted to study the very low
temperature properties and various details in the spectra. At finite
temperatures, the most vexing technical difficulty of the NRG is the
broadening problem discussed previously. Since the DMFT consists of
iteratively solving the effective impurity problem, the spectral
function artifacts at $\omega \sim T$ tend to amplify. This is
particularly bothersome for calculations of transport properties
\cite{Jarrell:1994ut,Jarrell:1995te,grenzenbach2006} where the
contribution to the integrals comes mostly from the frequency window
$\omega \in [-5T:5T]$. Another issue in the DMFT(NRG) approach is the
calculation of the self-energy as the ratio of two Green's functions
\cite{bulla1998,bulla2008}
\begin{equation}
\Sigma_{\sigma}(z) = \frac
{ \corr{ [d_\sigma, H_\mathrm{int}] ; d^\dag_{\sigma} }_z }
{ \corr{ d_\sigma; d^\dag_{\sigma} }_z },
\end{equation}
where $H_\mathrm{int}$ is the interaction part of the Hamiltonian,
while $d_\sigma$ is the impurity-orbital annihilation operator. The
main problem is that the causality requirement $\Im
\Sigma(\omega+i\delta) < 0$ can be (slightly) violated at low
temperatures around the Fermi level. Ironically, the difficulties
occur at low temperatures and on low-frequency scale, i.e., in the
regime which is commonly believed to the the forte of the NRG, while
it performs rather better than expected on intermediate and high
temperature and frequency scales.

In this work we explore an alternative approach to spectral function
calculations at finite temperatures. 
We propose to calculate the Green's functions on the discrete set of
the imaginary Matsubara frequencies and then (when necessary) compute
the Green's function for real-frequency arguments by analytic
continuation using the Pad\'e approximants. Switching to the imaginary
axis in the NRG may seem like a step in the wrong direction, but it is
motivated by the following observations: i) The NRG spectra are not
affected by large stochastic errors like the Monte Carlo data. The
fine details are therefore not lost and can be resolved by analytic
continuation. ii) One directly obtains both real and imaginary parts
of the Green's function on the imaginary axis, thus Kramers-Kronig
transformations are no longer necessary. iii) In the DMFT(NRG) loops,
most steps of the calculation can be performed in the Matsubara space;
an analytic continuation is only necessary to compute the
hybridization function on the real axis which is, however, immediately
integrated over to obtain the discretization coefficients. iv) There
is less arbitrariness compared to the broadening approach.

The paper is structured as follows. In Sec.~\ref{secmethod} we
describe the basic steps in the DMFT calculations, the NRG technique,
and our implementation of the Pad\'e approximation. In
Sec.~\ref{secsiam} we test the method on single-impurity problems: we
compute the temperature dependence of the spectral function of the
resonant-level model and the single-impurity Anderson model, and
compare various approaches. In Sec.~\ref{secdmft} we extend this
calculation with a self-consistency loop and study the Hubbard model
at finite temperatures. %
We conclude by discussing the relative merits of the different
techniques.

\section{Method}
\label{secmethod}

\subsection{Dynamical mean-field theory}

The DMFT is based on the observation that in the limit of infinite
dimensions $\Sigma(\omega,\vc{k}) \rightarrow \Sigma(\omega)$,
which implies the possibility of exactly mapping the lattice problem
onto a single impurity problem subject to self-consistency conditions
\cite{georges1992,georges1992mott,rozenberg1992,jarrell1992dmft}.
Let us consider the Hubbard model \cite{hubbard1963,kanamori1963,gutzwiller1963}
\begin{equation}
H_\mathrm{Hubbard}=\sum_{\vc{k}\sigma} (\epsilon_\vc{k}-\mu) c^\dag_{\vc{k}\sigma}
c_{\vc{k}\sigma} + U \sum_i n_{i\uparrow} n_{i\downarrow},
\end{equation}
describing a lattice of sites indexed by $i$ which can be
occupied by electrons with spin $\sigma=\uparrow$ and
$\sigma=\downarrow$. Here $\epsilon_\vc{k}$ is the non-interacting
band dispersion, $\mu$ is the chemical potential, and $U$ the Hubbard
repulsion. Furthermore, $c_{i\sigma}=1/\sqrt{N} \sum_\vc{k}
e^{i\vc{k}\cdot \vc{r}_i} c_{\vc{k}\sigma}$, and $n_{i\sigma} =
c^\dag_{i\sigma} c_{i\sigma}$ is the local occupancy.

In the DMFT, the information about the lattice structure is fully captured by
the non-interacting density of states (DOS)
\begin{equation}
\rho_0(\epsilon) = \frac{1}{N} \sum_\vc{k} \delta(\epsilon-\epsilon_\vc{k}).
\end{equation}
Using the Hilbert transform, one can compute the corresponding Green's
function $G_0(z)$ in the full complex plane:
\begin{equation}
G_0(z) = \int_{-\infty}^{\infty} \frac{\rho_0(\epsilon)\mathrm{d}\epsilon}{z-\epsilon}.
\end{equation}
For common lattices (such as Bethe lattice, 2D, 3D, and
infinite-dimensional cubic lattice, as well as for flat band) there
exist analytic closed-form expressions for $G_0(z)$.

The Hubbard model maps upon the single-impurity Anderson model (SIAM)
\cite{anderson1961,anderson1967,anderson1978,georges1992,georges1992mott,rozenberg1992,jarrell1992dmft,sakai1994,bulla1999}:
\begin{equation}
\begin{split}
H_\mathrm{SIAM} &=
\sum_{k\sigma} (\varepsilon_k-\mu) f^\dag_{k\sigma}
f_{k\sigma} - \mu n
+ U n_\uparrow n_\downarrow \\
&+ \sum_{k\sigma} V_k \left( f^\dag_{k\sigma} d_\sigma +
\text{H.c.} \right),
\end{split}
\end{equation}
where the operators $f_{k\sigma}$ annihilate an electron in the
continuum bath and $d_\sigma$ on the impurity site, while
$n=n_\uparrow+n_\downarrow$ with $n_\sigma=d^\dag_\sigma d_\sigma$.
The (complex) hybridization function $\Delta$, defined as
\begin{equation}
\Delta(z) = \sum_k \frac{|V_k|^2}{z-\varepsilon_k},
\end{equation}
fully characterizes the coupling between the impurity and the
continuum \cite{bulla1997}, i.e., there can be different (but
physically fully equivalent) descriptions in terms of the bath
energies $\varepsilon_k$ and hopping constants $V_k$.

The quantity of main interest is the momentum-resolved Green's
function 
\begin{equation}
G_{\vc{k}\sigma}(z) =
\corr{c_{\vc{k}\sigma};c^\dag_{\vc{k}\sigma}}_z
\end{equation}
and the corresponding
spectral function
\begin{equation}
A_{\vc{k}\sigma}(\omega)=-\frac{1}{\pi} \Im
G_{\vc{k}\sigma}(\omega+i\delta).
\end{equation}
They provide information about the electron dispersion in the presence
of interactions and can be used to determine the thermodynamic and
transport properties of the system within the linear response theory
\cite{Jarrell:1994ut,Jarrell:1995te,georges1996}. The output from
the impurity solver is the local impurity self-energy
$\Sigma_\sigma(z)$, which in the DMFT is taken to be equal to the
lattice self-energy of the correlated electron problem, thus
\begin{equation}
G_{\vc{k}\sigma}(z) = \frac{1}{z+\mu-\epsilon_\vc{k}-\Sigma_\sigma(z)}.
\end{equation}
The local ($\vc{k}$-averaged) lattice Green's function is then
\begin{equation}
\label{eq:loc}
\begin{split}
G_{\mathrm{loc},\sigma}(z) &= \frac{1}{N} \sum_\vc{k}
\frac{1}{z+\mu-\epsilon_\vc{k}-\Sigma_\sigma(z)} \\
&= \int \frac{\rho_0(\epsilon)\mathrm{d}\epsilon}{[z+\mu -\Sigma_\sigma(z)]-\epsilon} \\
&= G_0[z+\mu-\Sigma_\sigma(z)].
\end{split}
\end{equation}
The DMFT self-consistency condition relates the local lattice Green's
function and the hybridization function through
\begin{equation}
\Delta_\sigma(z)= 
z + \mu - \left(G^{-1}_{\mathrm{loc},\sigma}(z)+\Sigma_\sigma(z)\right).
\end{equation}
The hybridization function is then used as the input to the impurity
solver in the next step of the DMFT iteration. The calculation
proceeds until two consecutive solutions for the local spectral
function differ by no more than some chosen convergence criterium. The
approach to self-consistency can be significantly accelerated using
Broyden mixing, which can also be very efficiently used to control
the chemical potential $\mu$ in fixed occupancy calculations
\cite{broyden}.

\subsection{Numerical renormalization group}

Wilson's NRG is a non-perturbative numerical renormalization group
approach applied to quantum impurity problems \cite{wilson1975}. 
The cornerstone of the method is the {\it logarithmic discretization} of
the conduction band \cite{wilson1975,campo2005,resolution,odesolv}.
The infinite number of the continuum degrees of freedom is reduced to
a finite number, making the numerical computation tractable. The
discretization is chosen to be logarithmic because in the Kondo
problem the excitations from each energy scale contribute equally to
the renormalization of the effective exchange coupling
\cite{anderson1970}. The band is divided into slices of exponentially
decreasing width, 
\begin{equation}
\begin{split}
I_m^- &= [-\Lambda^{-m},-\Lambda^{-(m+1)}]D, \\
I_m^+ &= [\Lambda^{-(m+1)},\Lambda^{-m}]D
\end{split}
\end{equation}
for holes and electrons, respectively, with $m\geq 0$. Here $D$ is the
half-bandwidth, while $\Lambda>1$ is known as the {\it discretization
parameter}. A complete set of wave-functions is then constructed in
each interval $I_m^\pm$, the first chosen so that it couples to the
impurity, while other Fourier components are localized away from
it; only the first wave-function is retained, while all the others are
dropped from consideration \cite{wilson1975}. This approximation
becomes exact in the $\Lambda \to 1$ limit, but the accumulated
experience with the method indicates that it is reliable even for very
large $\Lambda$ despite the seeming crudeness of the approximation
\cite{campo2004}. The problem is then transformed to a tridiagonal
basis using the Lanczos algorithm, the result being a tight-binding
Hamiltonian which is known as the {\it hopping Hamiltonian} or the
{\it Wilson chain}. A number of improvements have been devised over
the years. On one hand, it has been shown that it is advantageous to
perform calculations for several interpenetrating discretization
meshes and then average the results \cite{yoshida1990,campo2005}. Such
$z$-{\it averaging} (or {\it twist-averaging}) allows more accurate
calculations at larger values of $\Lambda$ since the discretization
artifacts tend to largely cancel out for aptly chosen meshes. On the
other hand, modified discretization schemes further reduce some
systematic errors of the original Wilson's approach
\cite{campo2005,resolution,odesolv}. These improvements are crucial if
the goal is to obtain accurate results.

The NRG calculations in this work have been performed with the ``NRG
Ljubljana'' package \cite{nrglj}, which consists of two parts.
The first is a Mathematica program which initializes the problem by
performing the exact diagonalization of the initial Hamiltonian in a
chosen symmetry-adapted basis, and by transforming the operators of
interest to the eigenbasis. The calculations are performed using a
computer algebra system, and the input to the program are expressions
in the familiar second quantization notation \cite{sneg}. All
quantities are stored in the form of irreducible matrix elements, and
the Wigner-Eckart theorem is used to take into account the symmetry
properties. The second part of the package is the C++ code which
performs the iterative diagonalization. This consists of adding the
Wilson chain sites one by one, each time constructing the Hamiltonian
matrices in all invariant subspaces, diagonalizing them, and
transforming all the necessary operator matrices. The full description
at step $N$ of the iteration corresponds to the effective behavior of
the system on the temperature/energy scale $\omega_N \sim
\Lambda^{-N/2}$. Since the Fock space grows exponentially along the
chain, only a small part of the computed eigenstates are retained
after each step. A convenient way to perform this truncation is to
keep the states up to some suitable energy cutoff $E_\mathrm{cutoff}$
defined in terms of $\omega_N$. Alternatively, at most
$N_\mathrm{keep}$ states are retained. The systematic error introduced
by truncation is small due to the energy-scale separation property
\cite{wilson1975,errors} of the quantum impurity models: the matrix
elements between excitations of vastly different energy scales are
very small. The dynamical quantities are computed using the spectral
decomposition (Lehmann representation) for the eigenstates of the full
Hamiltonian. The original approach was based on the observation by
Sakai et al. \cite{sakai1989} that as one proceeds from one step to
the next, the lowest few eigenstates split due to the interaction with
the added shell states, while the intermediate lower levels do not
show any essential change. The intermediate states thus form a good
approximation of the eigenstates of the Hamiltonian in the
infinite-chain limit. For problems where the high-energy spectral
features depend on the low-energy behavior of the system, the spectral
function has to be computed taking into account the reduced density
matrix obtained from the density matrix of the low-energy fixed-point
\cite{hofstetter2000,sidecoupled,toth2008nrg,Weichselbaum:2012tq}. It
has been shown that a complete basis for the full Fock space of the
problem can be defined by judiciously using the information from the
discarded part of the NRG eigenstates \cite{anders2005,anders2006}.
This method does not suffer from over-counting of excitations and it
fulfills the normalization sum rule \cite{peters2006}. At finite
temperatures, this scheme can be improved by including the
contributions to the density matrix from all NRG shells
\cite{weichselbaum2007}. This full-density-matrix (FDM) approach is
currently the most reliable method for computing the
finite-temperature spectral functions using the NRG. In the DMFT
applications, one is particularly interested in the self-energy
$\Sigma$, which may be computed as the ratio \cite{bulla1998}
\begin{equation}
\label{eq:self}
\Sigma_\sigma(z) = \frac{F_\sigma(z)}{G_\sigma(z)},
\end{equation}
where
\begin{equation}
F_\sigma(z)=\corr{[d_\sigma,H_\mathrm{imp}]; d^\dag_\sigma}_z.
\end{equation}
This is known as the self-energy trick (or $\Sigma$-trick)
\cite{bulla1998}, and it has recently also been implemented in QMC
calculations \cite{2012PhRvB..85t5106H}.

Since the hopping Hamiltonian is finite, the computed raw spectral
functions are represented as a sum of delta peaks:
\begin{equation}
A_{\mathrm{NRG}, \sigma}(\omega) = \sum_j w_{j,\sigma}
\delta{}(\omega-\omega_{j,\sigma}).
\end{equation}
Alternatively, one may use fine-grained binning of these delta peaks
into very narrow intervals on a logarithmic grid (for instance 1000
bins per frequency decade).  To obtain a meaningful continuous
function, these peaks need to be broadened. The original approach to
obtaining a smooth curve was by Gaussian broadening followed by
separate spline interpolation of the results in odd and even steps, and
averaging of the two curves \cite{sakai1989}. A better
approach is broadening by the log-Gaussian
distribution function \cite{bulla2001}: each data point (delta peak at
$\omega_j$) is smoothed into
\begin{equation}
F(\omega,\omega_j) = \frac{e^{-b^2/4} \theta(\omega\omega_j)}{b\sqrt{\pi}}
\exp\left( - \frac{\ln^2|\omega/\omega_j|}{b^2} \right),
\label{ker1}
\end{equation}
i.e., a Gaussian function on the logarithmic scale, where $b$ is the
{\it broadening parameter}, chosen depending on the value of the
discretization parameter $\Lambda$ and the number of interleaved
discretization meshes $N_z$. Peaks sharper than the width of the
broadening kernel will appear broader than they truly are. Typically,
the value of $b$ is chosen to be $0.6$ or less, but with large $N_z$
and small $\Lambda$ it can be much reduced, to the point of largely
eliminating the NRG overbroadening problems (at the cost of much
longer computation time) \cite{resolution,errors}. At finite
temperatures, the following broadening kernel has been proposed
\cite{weichselbaum2007}:
\begin{equation}
K(\omega,\omega_j)=L(\omega,\omega_j)h(\omega_j)+G(\omega,\omega_j)[1-h(\omega_j)],
\end{equation}
where
\begin{equation}
\begin{split}
L(\omega,\omega_j) &= \frac{\theta(\omega\omega_j)}{\sqrt{\pi}\alpha
|\omega|} \exp \left[ -\left( \frac{\ln|\omega/\omega_j|}{\alpha}-\gamma
\right)^2 \right], \\
G(\omega,\omega_j) &= \frac{1}{\sqrt{\pi}\omega_0}
\exp\left[- \left( \frac{\omega-\omega_j}{\omega_0} \right)^2 \right], \\
h(\omega_j) &=
\left\{
\begin{array}{ll}
1, & |\omega_j|\geq \omega_0 \\
\exp\left[ -\left( \frac{\log|\omega_j/\omega_0|}{\alpha} \right)^2
\right], & |\omega_j|<\omega_0
\end{array} 
\right.\ .
\end{split}
\end{equation}
Here $\alpha$ is the broadening parameter for the log-Gaussian part,
equivalent to $b$ in the kernel in Eq.~\eqref{ker1}, $\gamma=\alpha/4$,
while $\omega_0$ is the cut-off where the log-Gaussian goes smoothly
into the Gaussian part; typically $\omega_0$ is chosen to be of the
order of the temperature $T$. This broadening approach leads to
sizable artifacts on the scale of $\omega_0$. We find that in practice
it is better to use slightly modified kernel:
\begin{equation}
K(\omega,\omega_j)=L(\omega,\omega_j)h(\omega)+G(\omega,\omega_j)[1-h(\omega)],
\end{equation}
which differs only in the argument of the cross-over function $h$.
This breaks the normalization condition, but produces smoother spectra
and the normalization is reestablished using the self-energy trick.
This procedure still leads to a slight bump around $\omega =
\omega_0$, which can be further smoothed out by averaging over several
choices of $\omega_0$. The artifacts can, however, never be completely
eliminated. In the following subsection we therefore discuss an
entirely different approach to obtaining a continuous spectral
function at finite temperatures.

\subsection{Pad\'e approximation}

The Green's function $G_{\sigma}(z)$ is
related to its corresponding spectral function by
\begin{equation}
\mathrm{Im}G_{\sigma}(\omega + i\delta{}) = -\pi{}A_{\sigma}(\omega)
\end{equation}
and similarly for $F_{\sigma}(z)$. %
Instead of performing the broadening, the complex functions
$G_{\sigma}(z)$ and $F_{\sigma}(z)$ (both analytic in the upper
complex half-plane, cf. Titchmarsh's theorem) are evaluated at the
Matsubara frequencies
\begin{equation}
i\omega_n = i(2n+1)\pi{}T
\end{equation}
via the Hilbert transform:
\begin{equation}
\label{eq:HilbertTransformG}
G_{\sigma}(i\omega_n) = \int_{-\infty}^{\infty}
\frac{A_{\mathrm{NRG},\sigma}(\omega)}{i\omega_n-\omega} \mathrm{d}\omega,
\end{equation}
and similarly for $F_{\sigma}(i\omega_n)$. Note that there is no need
to perform the Kramer-Kronig transformation to calculate the real parts of
$G_{\sigma}(\omega)$ and $F_{\sigma}(\omega)$. The equation
\eqref{eq:HilbertTransformG} is actually just a finite sum
\begin{equation}
G_{\sigma}(i\omega_n) = \sum_j
\frac{w_{\sigma,j}}{i\omega_n-\omega_{\sigma,j}}.
\end{equation}
If the $z$-averaing is used, it is performed at this point:
\begin{equation}
G_{\sigma}(i\omega_n) = \frac{1}{N_z} \sum_{n=1}^{N_z} \sum_j
\frac{w^n_{\sigma,j}}{i\omega_n-\omega^n_{\sigma,j}}.
\end{equation}
The self-energy is calculated on the imaginary axis in the same way as
on the real axis via Eq.~\eqref{eq:self} which holds in the whole
upper complex half-plane.  
From the self energy in the Matsubara frequencies
$\Sigma_{\sigma}(i\omega_n)$, one calculates the local Green's
function $G_{\mathrm{loc},\sigma}(i\omega_n)$ using
Eq.~\eqref{eq:loc}. To complete the DMFT loop, one needs to calculate
the hybridization function on the real axis. We first calculate
\begin{equation}
\Delta_{\sigma} (i\omega_n) = i\omega_n + \mu - \left(
G_{\mathrm{loc},\sigma}^{-1}(i\omega_n) - \Sigma_{\sigma}(i\omega_n) \right). 
\end{equation}
This function is analytic in the upper complex half plane so we can
perform an analytic continuation and evaluate $\Delta$ just above the real axis to
obtain the (real) hybridization function $\Gamma_\sigma(\omega)$:
\begin{equation}
\label{eq:HybrdiMats}
\Gamma_{\sigma}(\omega) = -\mathrm{Im} \left[ \Delta_{\sigma}(\omega + i\delta)
\right].
\end{equation}
An analytic continuation is also needed to determine the spectral
function 
\begin{equation}
\label{eq:localSpectral}
A_{\sigma}(\omega) = -\frac{1}{\pi} \mathrm{Im} \left[ G_{\mathrm{loc},\sigma}(\omega
+ i\delta) \right].
\end{equation}

The generalized mathematical problem of analytic continuation can be
stated as follows: find an analytic function $f(z)$ in the upper
complex half plane that coincides with calculated values on the discrete
set of points 
\begin{equation}
\label{eq:PadeSystem}
\{ f(z_j)=f_j \},
\end{equation}
where $(z_j, f_j)$ are the known point-value pairs of the function. The
function $f(z)$ must also obey the  asymptotic behavior of the Green's
function [or $\Delta_{\sigma}(z)$], namely 
\begin{equation}
f(z) \sim \frac{1}{z}.
\end{equation}
There exist two main numerical techniques for the analytic
continuation: the maximum entropy method \cite{PhysRevB.41.2380} (MEM)
and Pad\'e approximation \cite{PhysRevB.61.5147}. MEM is essentially
an improved fit to available data, taking into account known physical
properties of the Green's/spectral function, such as sum rules,
possible symmetries and higher moments of the distribution. 
In this article, we focus on Pad\'e approximation; the comparison of
MEM and Pad\'e in the context of NRG are yet to be explored.

The Pad\'e approximation method is based on the assumption that $f(z)$
is a rational function
\begin{equation}
\label{eq:PadeDef}
f(z) = \frac{p_0+p_1z+\cdots + p_rz^r}{q_0+q_1z+\cdots{}+q_rz^r +
z^{r+1}},
\end{equation}
where $p_j$ and $q_j$ are the unknown complex coefficients to be
determined. Inserting the
Pad\'e approximant \eqref{eq:PadeDef} into equations
\eqref{eq:PadeSystem} for each point $(z_j, f_j)$ generates a linear
system of equations. Defining a vector of unknowns 
\begin{equation}
\bm{x} = [p_0,
p_1, {\ldots} , p_r, q_0, q_1, {\ldots} , q_r ],
\end{equation}
a right-hand-side vector 
\begin{equation}
\bm{b}=[f_0z_0^r, f_1z_1^r, {\ldots} , f_{2r-1}z_{2r-1}^r],
\end{equation}
and a matrix
\begin{equation}
\label{eq:PadeMatrix_a} 
\bm{A} =
\left( \begin{array}{cccccccc}
1 & z_0 & z_0^2 & \cdots{} & f_0 & f_0z_0 & \cdots{}  \\
1 & z_1 & z_1^2 & \cdots{} & f_1 & f_1z_1 & \cdots{}  \\
\vdots & \vdots & \vdots & \vdots & \vdots & \vdots & \vdots
\end{array} \right),
\end{equation}
the solution for the coefficients is
\begin{equation}
\bm{x}=\bm{A}^{-1}\bm{b}.
\end{equation}
We have assumed that the number of points $(z_j, f_j)$ is $2r$. Alternatively,
one can use recursive relations of the continued fractions representation of the
Pad\'e approximant to evaluate it at specific points \cite{Vidberg:1977vo}.

Fitting a rational function is a numerically ill-defined problem. Let us
define $\xi$ as the ratio between the biggest and the smallest element
in the matrix $\bm{A}$. If $z_j$ are the Matsubara frequencies, the
ratio is approximately
\begin{equation}
\xi = {\left[ (4r+1)\pi{}T  \right] }^{\pm{}r},
\end{equation}
where we take the minus sign in the power if the base is smaller than
$1$. In order to invert the matrix $A$, the numeric precision of
$2\log_{2} \xi$ binary digits is needed \cite{PhysRevB.61.5147},
i.e., much more than the standard $53$-bit mantissa of $64$-bit
double-precision floating point numbers. For this reason, in our
implementation of the Pad\'e method we use the GNU Multiple Precision
Arithmetic Library (GMP) for arbitrary precision floating-point
numerics. To solve the linear system of equations, we perform Gaussian
elimination without pivoting. The process can be performed in $O(r^3)$
multiplications. Each multiplication takes $O(b \log{b})$ CPU cycles,
where $b$ is the number of mantissa bits in the floating point used
(using fast Fourier transform multiplication). 

When the coefficients of the Pad\'e approximant are known to high
precision, one can approximately calculate the value of the Green's
function in any point of the upper complex half plane using Horner's 
polynomial evaluation scheme. The most
interesting are the values just above the real axis to compute the
hybridization function $\Gamma_{\sigma}(\omega)$ via
Eq.~\eqref{eq:HybrdiMats} or the spectral function
$A_{\sigma}(\omega)$ via Eq.~\eqref{eq:localSpectral}. The Pad\'e
approximant can be, however, also used for extrapolation when an
insufficient number of Matsubara frequencies have been computed. 
It provides a good fit to tails even when the asymptotic behavior is
not yet reached and simple asymptotic $1/z$ fit on tails does not work. 

In most calculations we use $N_m=2r=350$ Matsubara frequencies and
internal matrix inversion precision of $1024$ mantissa floating bits.
In most cases, taking more points does not improve the solution; in
fact, sometimes taking less points helps avoid some issues. The
precision of input points plays a big role in how good the Pad\'e
approximant is, as already discussed in
Ref.~\onlinecite{PhysRevB.61.5147}. 

We remark that one can also determine the coefficients of the rational
function of order $r$ using the data from more than $N_m=2r$ points by
least-squares fitting.

\section{Results for a  single-impurity problem: Anderson model}
\label{secsiam}

We first consider a single impurity in a metal host described by the
SIAM, now written as
\begin{equation}
\begin{split}
H_\mathrm{SIAM} &=
\sum_{k\sigma} \varepsilon_k f^\dag_{k\sigma}
f_{k\sigma} + \epsilon n
+ U n_\uparrow n_\downarrow \\
&+ \sum_{k\sigma} V_k \left( f^\dag_{k\sigma} d_\sigma +
\text{H.c.} \right),
\end{split}
\end{equation}
where $\epsilon$ is the impurity level, while the chemical potential
is fixed to zero in this section, $\mu=0$. The spectral function
can in general be expressed as
\begin{equation}
\label{A}
A(\omega) = -\frac{1}{\pi} \Im \left( \frac{1}{\omega+i\delta -\epsilon-
\Sigma(\omega)-\Delta(\omega)} \right).
\end{equation}
When the continuum is modelled as a flat band, i.e., a band with a
constant density of states $\rho$ in the interval $\omega \in [-D:D]$
which couples to the impurity with a constant hopping $V_k \equiv V$ so that the
hybridization strength 
\begin{equation}
\Gamma=\pi\rho V^2
\end{equation}
is a constant, the complex hybridization function $\Delta(z)$ is given
as
\begin{equation}
\Delta(z) = -\frac{\Gamma}{\pi} \ln\left( \frac{z/D-1}{z/D+1} \right),
\end{equation}
or, on the real-frequency axis for $-D < \omega <D$,
\begin{equation}
\Delta(\omega+i\delta) = -\Gamma \left[ i + \frac{1}{\pi}
\ln\left( \frac{1-\omega/D}{1+\omega/D} \right) \right].
\end{equation}
In the limit $U=0$, the SIAM is equal to the non-interacting
resonant-level model whose spectral function is given exactly by
Eq.~\eqref{A} with $\Sigma(\omega) \equiv 0$. For
small $\Gamma$, the spectral function is well approximated by a
Lorentzian
\begin{equation}
A(\omega) = \frac{\Gamma/\pi}{(\omega-\epsilon)^2+\Gamma^2}.
\end{equation}
For larger $\Gamma$, the hybridization self-energy effects become
sizable, thus the curve shape deviates from a pure Lorentzian form,
the peak is shifted to
\begin{equation}
\omega_0 = \epsilon-\frac{\Gamma}{\pi}\ln\left( \frac{1-\omega_0/D}
{1+\omega_0/D} \right),
\end{equation}
and, in addition, near the band edges $\omega = \pm D$ the spectral
function shows additional features: local minima at
\begin{equation}
\omega_\mathrm{min} = \pm D \sqrt{1-\frac{2\Gamma}{\pi D}},
\end{equation}
and local maxima at $\omega_\mathrm{max}$ given as the solutions of the
equation
\begin{equation}
\pi(\omega_\mathrm{max}-\epsilon)+\Gamma \ln\left(
\frac{1-\omega_\mathrm{max}/D}{1+\omega_\mathrm{max}/D} \right)=0,
\end{equation}
located near the band edges. For $\epsilon=0$ and moderately large
$\Gamma$, they are approximately equal to
\begin{equation}
\omega_\mathrm{max} = \pm \left[ 1- \exp\left( -\frac{\pi D}{\Gamma}
\right) \right],
\end{equation}
i.e., the maxima are located exponentially close to the band edges.
Beyond $\omega_\mathrm{max}$ the spectral function goes to zero at
$\omega = \pm D$ in a continuous way, but with a diverging slope.
Similar behavior is expected in all models where the imaginary part of
the hybridization function drops to zero discontinuously, so that its
real part (Hilbert transform) has a logarithmic singularity. 
This is the case, for example, for a 2D cubic lattice DOS with
\begin{equation}
\Delta(z)=\frac{2\Gamma D}{\pi^2 z} K(D^2/z^2),
\end{equation}
where $K$ is the elliptic function. The weight of the band-edge
features is exponentially small for small $\Gamma$, but may be
appreciable for moderate $\Gamma$ of the order of the bandwidth, or if
the main spectral peak is close to the band edge (for instance, a few
times $\Gamma$). Such features typically are not visible in the NRG
results because of the overbroadening effects, although in a careful
calculation with small broadening width and for many values of the
twist parameter $z$, they may be observed \cite{resolution}.  In the
other case, when $\Im\Delta$ goes to zero continuously, the spectral
function decreases monotonously near the band edges. We note that in
non-interacting models, the spectral function does not depend on the
temperature.

The NRG calculations in this section are performed with the
discretization parameter $\Lambda=2$, with the $z$-averaging over
$N_z=16$ values of the twist parameter using a modified discretization
scheme which eliminates the band-edge artifacts of the conventional
approach \cite{resolution,odesolv}. Spectral functions are computed
using the FDM NRG scheme and the $\Sigma$-trick. The truncation cutoff
is set at $E_\mathrm{cutoff}=10\omega_N$ unless otherwise noted. Flat
band with constant $\Gamma$ is used in all cases.

\subsection{Non-interacting case, $U=0$}

\begin{figure}[htbp]
\centering
\includegraphics[clip,width=0.48\textwidth]{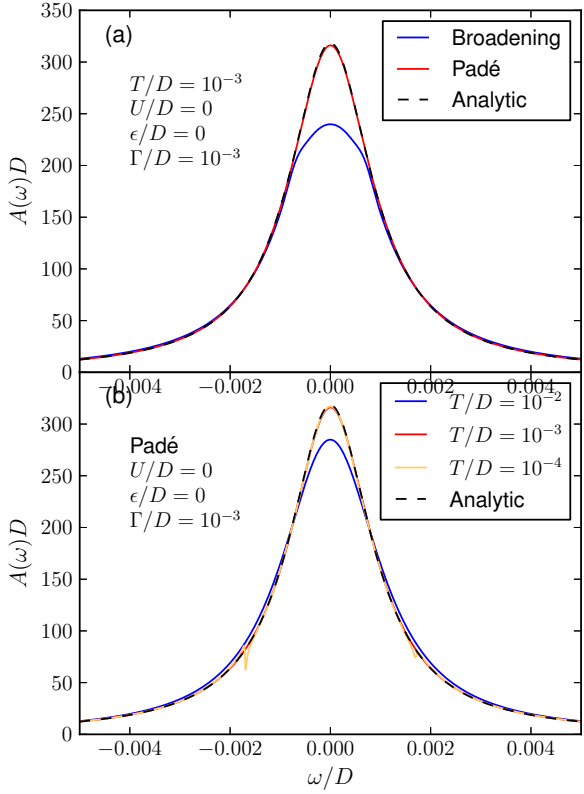}
\caption{(Color online) (a) Spectral function of the non-interacting
resonant-level model obtained with broadening ($\alpha=0.2$) and with
the Pad\'e approximant. The Pad\'e approximant overlaps nearly
perfectly with the analytic result. (b) Spectral functions calculated
at different temperatures $T$ using the Pad\'e approximant (i.e.,
using the sets of Matsubara points $i\omega_n=i(2n+1)\pi T$ with the raw
results of NRG calculations performed at chosen temperatures $T$). The
analytic result does not depend on the temperature.
Deviations in the numerical results occur for $T \gtrsim \Gamma$.
}
\label{fig:1}
\end{figure}

\begin{figure}[htbp]
\centering
\includegraphics[clip,width=0.48\textwidth]{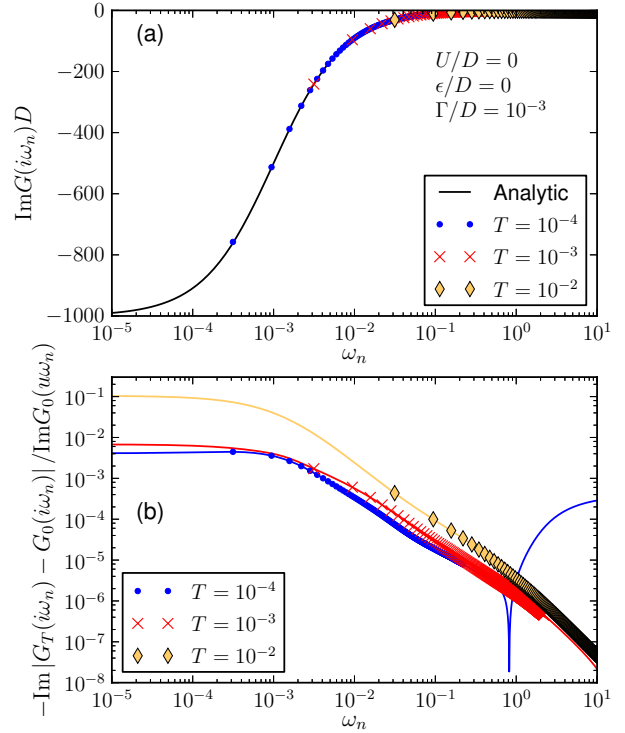}
\caption{(Color online) (a) Imaginary part of the Green's function for
the non-interacting resonant-level model (same parameters as in
Fig.~\ref{fig:1}) evaluated at the Matsubara frequencies for different
temperatures. The analytic solution is shown with solid black line.
(b) Relative errors of raw Matsubara spectral functions (dots) and the
corresponding Pad\'e approximants (solid lines).
The dots quantify the intrinsic NRG errors, while the lines measure
the full systematic error (intrinsic + analytic continuation errors). 
}
\label{fig:greenMatsubara_vs_T}
\end{figure}

Numerical calculation of a spectral function using the NRG is a
non-trivial test of the method even in the absence of interactions. 
Furthermore, even though the spectral function of a non-interacting
problem does not depend on the temperature, numerical approaches
generally have more or less severe difficulties reproducing this
simple fact. In Fig.~\ref{fig:1} we therefore first analyze the NRG
results for the non-interacting resonant-level model obtained by the
conventional broadening technique with a relatively small broadening
parameter $\alpha=0.2$ and by the proposed Pad\'e approximant method,
and compare them with the exact result. The $\Sigma$-trick is
obviously not used here, because it would trivially produce the exact
result. We consider a peak centered at $\omega=\epsilon=0$, i.e., at
the Fermi level, where the NRG is said to have very good spectral
resolution. In panel (a) we clearly observe the advantages of the
Pad\'e procedure at finite temperatures (here $T=\Gamma$): the Pad\'e
approximant produces a spectral function which practically overlaps
with the analytic formula, unlike the broadening procedure which
produces a severely distorted spectral peak with missing spectral
weight at low frequencies (the spectral peak tails are, however, well
reproduced in both approaches; furthermore, we note that in the $T \ll
\Gamma$ limit, the broadening reproduces the exact results to a very
good approximation on all frequency scales, thus the problems become
manifest only at finite temperatures). In panel (b) we compare Pad\'e
approximant results obtained at different temperatures. The best
agreement with the exact results is obtained for $T \lesssim \Gamma$,
which is expected since for a temperature scale comparable to the
characteristic physical scales of the problem, the finite set of the
Matsubara point provides a well matched sampling. For $T \ll \Gamma$,
the agreement remains good but requires a sufficient number of the
Matsubara points $N_m$ (this issue is discussed in the following).
Furthermore, for $T=10^{-4}$ we observe some artifacts on the flanks
of the curve (these are also discussed later on). For $T \gg \Gamma$,
we find that the spectral height is underestimated also in the Pad\'e
approximant approach, but less severely than in the broadening scheme.
This actually indicates a limitation of the NRG method itself, not of
the Pad\'e approximation scheme. Even at high $T$ the Green's function
at the corresponding Matsubara frequencies should contain the
necessary information to reconstruct the Lorentzian peak, but the
output from the NRG itself already has sizable systematic errors. We
explore this question in more detail by plotting the raw Green's
function on the Matsubara axis in panel (a) of
Fig.~\ref{fig:greenMatsubara_vs_T}, as well as the relative error in
panel (b). The exact expression on the imaginary axis is (for
$\epsilon=0$ and flat band):
\begin{equation}
G(iy)= -\frac{i}{y+\frac{\Gamma}{\pi}\arg\left(\frac{iy-D}{iy+D}\right)}.
\end{equation}
We find that the systematic error of NRG is below $0.5\%$. 

\begin{figure}[htbp]
\centering
\includegraphics[clip,width=0.48\textwidth]{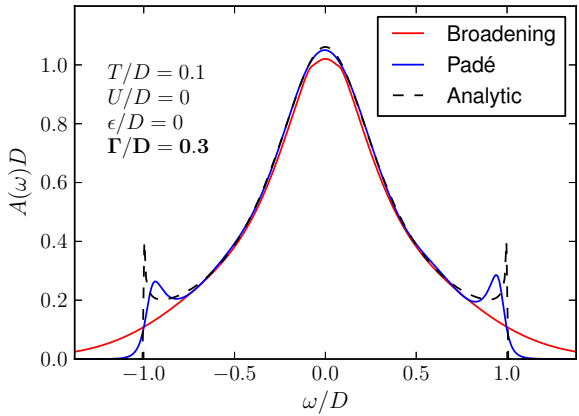}
\caption{(Color online) Spectral function of the non-interacting
resonant-level model with large hybridization $\Gamma/D=0.3$, where
the behavior at the band edges $\omega=-D$ and $\omega=D$ may be
compared. }
\label{fig:spectralPadeU0_vs_Gamma2}
\end{figure}

We study the behavior near the band edges using the resonant-level
model with a very large hybridization $\Gamma/D=0.3$, see
Fig.~\ref{fig:spectralPadeU0_vs_Gamma2}. The band-edge peaks are
completely washed out in the broadening approach and the spectral
function has long tails in the region where it should be equal to
zero. The Pad\'e approach resolves the band edges, although it is
unable to correctly resolve the shape and the width of the
near-band-edge resonances. Since the Pad\'e approximant is a rational
function, it cannot describe a precipitous drop to zero at band
edges. The Pad\'e approach also better describes the behavior at the
Fermi level, which is underestimated when broadening is used.

\begin{figure}[htbp]
\centering
\includegraphics[clip,width=0.48\textwidth]{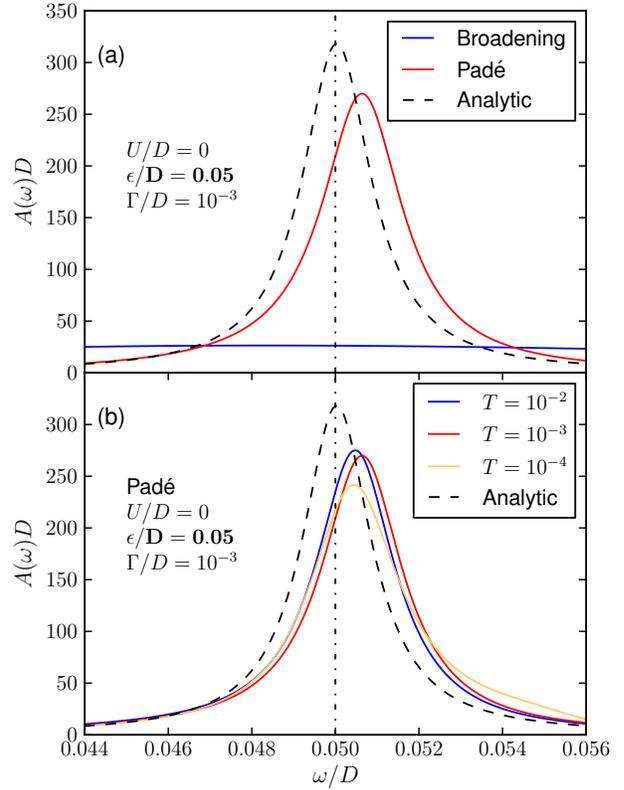}
\caption{(Color online) (a) Spectral function of the non-interacting
resonant-level model with a peak centered far away from the Fermi
level, $\epsilon \gg \Gamma$. Standard approach produces severely
overbroadened result since the kernel width for $\alpha=0.2$ is
$\Gamma_\mathrm{br}\approx \alpha \epsilon = 10^{-2}$. The
Pad\'e method is much more successful, although the peak is slightly
shifted away from its correct position. (b) Comparison of the spectral
functions computed with the Pad\'e approximant for a range of
temperatures. Similar degree of agreement is produced for all $T$
considered.}
\label{fig:spectralPadeU0eps0.05_vs_Gamma}
\end{figure}

We now consider the asymmetric case where the spectral peak is located
at $\epsilon \neq 0$, i.e., away from the Fermi level. This is the
situation where the broadening procedure has notorious difficulties.
The width of the log-Gaussian kernel is namely proportional to the
frequency, 
\begin{equation}
\Gamma_\mathrm{br} \approx \alpha \omega \approx \alpha
\epsilon,
\end{equation}
where $\alpha$ is the broadening parameter. If the intrinsic width of
the spectral feature considered is much less than
$\Gamma_\mathrm{br}$, significant overbroadening will occur. In the
example presented in Fig.~\ref{fig:spectralPadeU0eps0.05_vs_Gamma} we
take a rather extreme case of $\Gamma/D=10^{-3}$, while
$\Gamma_\mathrm{br}/D \approx 10^{-2}$ for $\alpha=0.2$ and
$\epsilon=0.05$. The plot in panel (a) indeed shows that the
broadening procedure produces a severely overbroadened peak, which is 
nearly a flat line on the scale of the figure. The Pad\'e approximant,
on the other hand, does not have difficulties reproducing narrow
Lorentzian peaks, as expected. We find, however, that the peak is
somewhat displaced from the correct position, and is slightly wider.
In panel (b) we show that the spectral functions depend on the
temperature in a non-systematic way, although the result is acceptable
for all temperatures considered. The deviations from the exact results
may again be related to the raw output from NRG, rather than to the
Pad\'e procedure.

The spectral function computed using Pad\'e approximation is not
guaranteed to fulfill the normalization sum rule. For peak centered at
the Fermi level, the normalization deviates from 1 to at most a few
per mil. For a peak away from the Fermi level, the deviations can be
sizable, up to several per cent, especially for larger $\Gamma$. In
principle, one could reformulate the Pad\'e approximation as a
least-squares-fitting procedure and implement the normalization sum
rule (and perhaps additional higher-moment sum rules) as a constraint
on the parameters. This is an idea worth pursuing.

\subsection{Interacting case, $U>0$}

\begin{figure}[htbp]
\centering
\includegraphics[clip,width=0.48\textwidth]{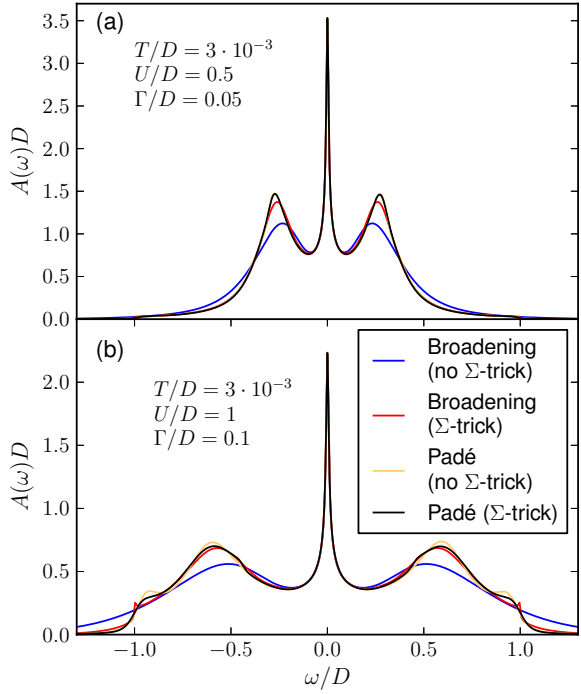}
\caption{(Color online) Spectral functions for the interacting
single-impurity Anderson model (SIAM) obtained with broadening
($\alpha=0.2$) and by Pad\'e approximation, with and without the
$\Sigma$-trick.  Two sets of parameters with the same $\Gamma/U$ ratio
are used, thus the effective Kondo exchange coupling constant $J$ is the
same. For $U/D=1$, the Hubbard peaks are located close to the band
edges, and additional spectral features can be resolved at the edge.
For $U/D=0.5$ the Pad\'e results with and without the $\Sigma$-trick
overlap nearly perfectly, while for $U/D=1$ there are some small
differences.
}
\label{fig:spectralBroadening_vs_PadeHalf}
\end{figure}

We now turn to the interacting case with finite $U$. For $U > \pi
\Gamma$ and near half-filling, the system is in the Kondo regime
\cite{anderson1961,anderson1967,haldane1977}. As the temperature is
reduced, on the temperature scale of $U$ the charge fluctuations on
the impurity site freeze out and the impurity may be described solely
in terms of its spin degrees of freedom
\cite{schrieffer1966,haldane1978,krishna1980a}; using the
Schrieffer-Wolff transformation \cite{schrieffer1966}, the model maps
onto the Kondo impurity model
\begin{equation}
H_\mathrm{Kondo} = \sum_{\vc{k}\sigma} \varepsilon_{k}
f^\dag_{\vc{k}\sigma}f_{\vc{k}\sigma} + J
\vc{S} \cdot \vc{s},
\end{equation}
where the Kondo exchange coupling constant $J$ is given by 
\begin{equation}
\rho J = \frac{2\Gamma/\pi}{-\epsilon} + \frac{2\Gamma/\pi}{\epsilon+U},
\end{equation}
$\vc{S}$ is the impurity spin operator, while $\vc{s}$ is the
conduction-band spin density at the impurity position.  At the Kondo
temperature, given by \cite{krishna1980a}
\begin{equation}
T_K \sim U \exp \left( -\frac{1}{\rho J} \right),
\end{equation}
the spin degree of freedom is screened by the conduction-band
electrons (the Kondo effect). The two characteristic scales are also
reflected in the spectral features: the charge fluctuations are
associated with spectral peaks at $\omega=\epsilon$ and
$\omega=\epsilon+U$ with half-width at half-maximum of $2\Gamma$,
while the screening leads to a Kondo resonance \cite{suhl1965} pinned
to the Fermi level, with a spectral width of the order of the Kondo
temperature. The zero-temperature spectral function at the
Fermi level is determined by the Friedel sum rule \cite{langreth1966}
\begin{equation}
A(0)=\frac{\sin^2 \delta_{\mathrm{q.p.}} }{\pi \Gamma},
\end{equation}
where $\delta_\mathrm{q.p.}$ is the quasiparticle scattering phase
shift in the local FL theory of the SIAM. If the system is
particle-hole (p-h) symmetric, i.e., if $\delta=\epsilon+U/2=0$, then
$\delta_\mathrm{q.p.} \equiv \pi/2$, irrespective of the $U/\Gamma$
ratio, thus there is a constraint $\pi \Gamma A(0)=1$. In the deep
Kondo regime, $U/\pi\Gamma \gg 1$, the phase shift is close to $\pi/2$
even if the p-h symmetry is slightly broken. Unless $\Gamma \ll D$, we
also expect some features near the band edges, just like in the
non-interacting case, as discussed above. At finite temperatures, the
Kondo resonance is washed out starting at $T \sim T_K$. In this work,
we use the definition of $T_K$ based on transport properties, i.e.,
$T_K$ is the temperature where the conductance through a nanodevice
described by the SIAM is reduced to one half the conductance quantum.
The conductance and the definition of $T_K$ are discussed more
thoroughly in Sec.~\ref{TKsec}.

In Fig.~\ref{fig:spectralBroadening_vs_PadeHalf} we plot the spectral
functions in the p-h symmetric case for two different choices of $U$
and $\Gamma$, but with the same $\Gamma/U$ ratio. The curves obtained
by broadening without the $\Sigma$-trick are rather typical of
standard NRG results: the Kondo resonance is well captured, but the
Hubbard peaks are significantly over-broadened. For $U/D=1$, the
spectral density remains finite even far outside the conduction band
due to the long tails of the broadening kernel. Furthermore, no
features near the band edges are detected. 

The curves obtained by the Pad\'e approach are significantly improved
even without the $\Sigma$-trick. They show all the expected features: the
Kondo resonance, the Hubbard peaks, and (for $U/D=1$) the
near-band-edge resonances, followed by a fast decay to zero
outside the band.

When the $\Sigma$-trick is used, the improvement is dramatic in the
broadening approach, since the over-broadening is strongly reduced.
For the $U/D=1$ case, the behavior near band edges is also improved. 
In the Pad\'e approach, however, the use of the $\Sigma$-trick has
very little effect. We thus conclude that {\it when the $\Sigma$-trick
is not used, the Pad\'e approximant approach vastly outperforms
broadening}, and that {\it the Pad\'e approach significantly reduces
the need for using the self-energy trick}. As discussed previously,
the $\Sigma$-trick is still very useful to restore the normalization of the
spectral function to 1.

\begin{figure}[htbp]
\centering
\includegraphics[clip,width=0.48\textwidth]{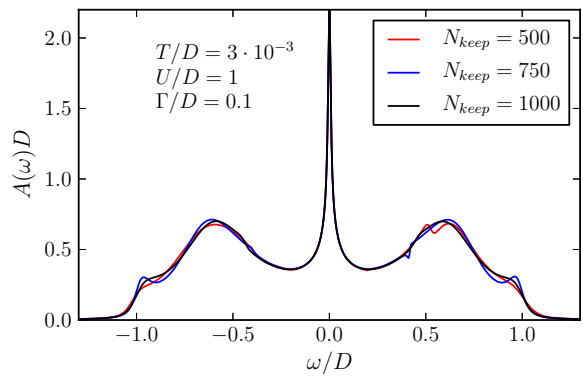}
\caption{(Color online) Spectral function of the single-impurity
Anderson model for different numbers of kept states,
$N_{\mathrm{keep}}$,
in the NRG procedure. The artifacts (for instance the dip at
$\omega/D=0.5$ for $N_\mathrm{keep}=500$) are truncation-cutoff
dependent, thus they stem from the raw NRG results.
}
\label{fig:spectralPade_vs_Keep}
\end{figure}

However, when using the Pad\'e approximant we also observe some anomalies.
Because they appear at different locations for different NRG
truncation parameters, see Fig.~\ref{fig:spectralPade_vs_Keep}, we
argue that they are a direct manifestation of the systematic NRG
errors that are reproduced too accurately by the Pad\'e approximation.
The standard broadening approach hides such anomalies, but they are,
in fact, present in the spectra and may be revealed in
``high-resolution'' calculations with small broadening-kernel width.
The Pad\'e approximation, however, appears to over-emphasize them.
Improved results may be obtained by averaging over
different truncation cutoffs \cite{errors}.

\begin{figure}[htbp]
\centering
\includegraphics[clip,width=0.48\textwidth]{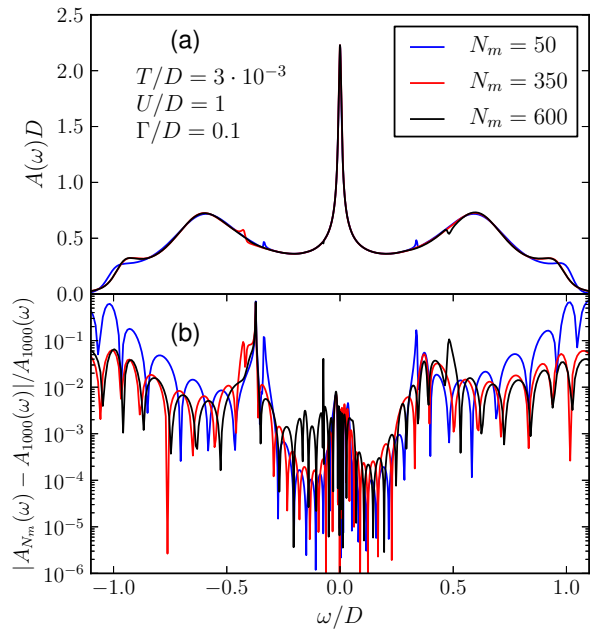}
\caption{(Color online) (a) Spectral functions of the
single-impurity Anderson model  obtained  with Pad\'e approximant using the first
$N_{m}$ Matsubara frequencies. (b) Relative difference between the
spectral functions computed with $N_m=1000$ Matsubara frequencies
compared to those for lower $N_{m}$.
}
\label{fig:spectralPade_vs_nrmats}
\end{figure}

In the broadening approach, there is significant arbitrariness in the
choice of the shape of the kernel, its width, and the approach to
handle the $\omega \lesssim T$ and $\omega \gtrsim T$ parts in
distinct ways. In the analytic continuation approach, there is in
principle no arbitrariness, since the information about the Green's
function on the set of Matsubara points fully and uniquely determines
the Green's function in the whole complex half-plane, in particular on
the real-axis. The way the continuation is performed is, clearly,
non-unique, but once the Pad\'e approximation approach is chosen, the
only adjustable parameter is the number of the Matsubara points taken
into account in the fitting procedure. 
In QMC calculations, it is practice to use sufficiently many Matsubara
frequencies to reach the asymptotic $1/z$ behavior on the imaginary
axis which corresponds to choosing $\omega_{N_{m}} \gg U$. (Note that
the number of points is temperature dependent.) In NRG calculation,
however, we observe that we can recover the tails using less
frequencies. The central peak is well reproduced using only $N_{m}=50$
frequencies and $N_{m}=350$ is enough to obtain essentially fully
converged spectral function, as shown in
Fig.~\ref{fig:spectralPade_vs_nrmats}(a). In
Fig.~\ref{fig:spectralPade_vs_nrmats}(b) we plot the relative
differences between the Pad\'e approximants for different numbers of
Matsubara points. In the Kondo peak region the differences are below
$0.01$, thus the relative error contributed by the analytic
continuation itself may be estimated to be well below one percent.
From this one may draw the conclusion that the main source of error at
low frequencies in this approach is the systematic error of the NRG
itself, not the analytic continuation procedure. At higher frequencies
the errors are slightly larger; in particular, at the artifacts
discussed above the error may locally be up to 10\%.

\begin{figure}[htbp]
\centering
\includegraphics[clip,width=0.48\textwidth]{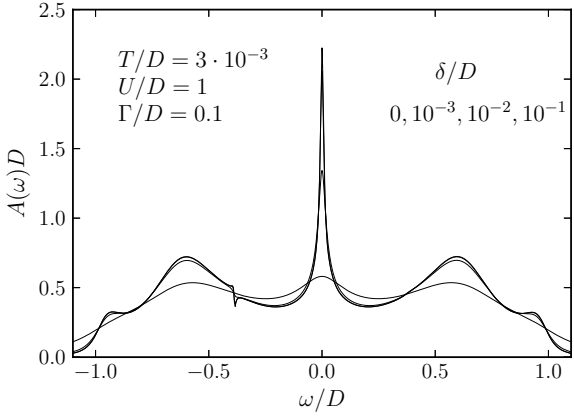}
\caption{Spectral functions of the single-impurity Anderson model
obtained with Pad\'e approach using different shifts from the real
axis.
}
\label{fig:spectralPade_vs_nmats}
\end{figure}

We find that the shift away from the real axis $\delta$ in the
expression for the spectral function $A(\omega)=-(1/\pi)\Im
G(\omega+i\delta)$ does not play an important role, since typically
there are no poles of the Pad\'e rational function on the real axis
(in which case it is, in fact, safe to put $\delta=0$). For large
$\delta$, artificial broadening is introduced, see
Fig.~\ref{fig:spectralPade_vs_nmats}.

\begin{figure}[htbp]
\centering
\includegraphics[clip,width=0.48\textwidth]{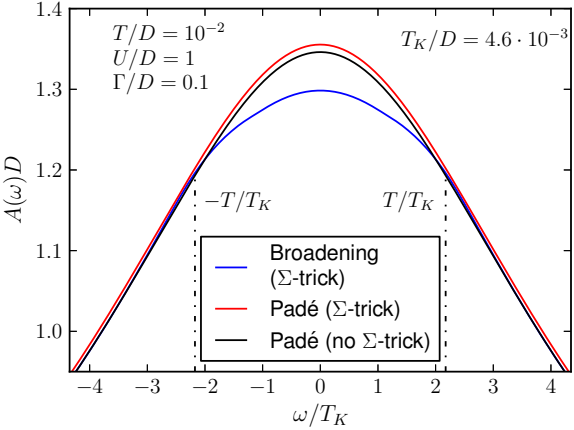}
\caption{(Color online) Close-up view of the top of the Kondo
resonance at a temperature greater than $T_K$. Note the deviation of
the peak shape from parabolic behavior when broadening is used.
}
\label{fig:spectralBroadening_vs_Pade_zoomPeakHigherT}
\end{figure}

Of particular interest is the low-energy part of the spectral
function, i.e., the Kondo resonance, because it is directly
responsible for the conductance anomalies observed in magnetically
doped metals and in semiconductor quantum dots. At non-zero
temperatures, the shape of the resonance is difficult to reliably
establish due to the broadening problems. Different version of the
broadening procedure yield improved results in some cases, but worse
in others. For example, an ``optimized'' scheme which produces less
artifacts when there is a low-frequency spectral peak may lead to more
pronounced artifacts in the case of a low-frequency spectral dip (and
vice versa). There is thus no universal choice. The comparison in
Fig.~\ref{fig:spectralBroadening_vs_Pade_zoomPeakHigherT}, indicates
that the Pad\'e approximant approach does not create bumps near
$\omega=\pm{}T$, unlike the broadening. In addition, it appears that
broadening leads to a significant underestimation of the height of the
spectral peak at its center. Using the $\Sigma$-trick in the Pad\'e
approach fixes the normalization problem.
Figure~\ref{fig:spectralBroadening_vs_Pade_zoomPeakHigherT}
illustrates one of the main results of this work: the Pad\'e
approximant approach allows to determine the finite-temperature
spectral functions with {\it less artifacts} on the scale of $\omega
\sim T$.

\begin{figure}[htbp]
\centering
\includegraphics[clip,width=0.48\textwidth]{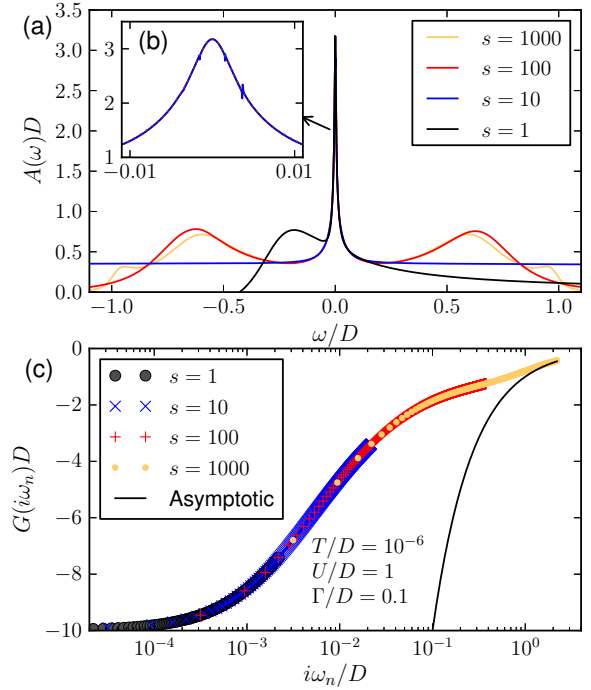}
\caption{(Color online) (a) Spectral functions of the SIAM at very low
temperature obtained using Pad\'e approximations which use
non-consecutive Matsubara frequencies as input: $i\omega_n=i(2n+1)\pi
s T$. The parameter $s$ quantifies the step length. The inset (b) shows a
close-up on the Kondo resonance. (c) Green's function on the
imaginary-frequency axis, indicating the Matsubara points taken into
account in the Pad\'e approximant construction. The black line is the
asymptotic $1/z$ form.}
\label{fig:spectralPade_vs_decimation}
\end{figure}

A function, analytic in the upper complex half-plane, is fully
determined by its values on a countably infinite set of points. When
fitting the Pad\'e approximant, one commonly takes the values of the
function at the Matsubara points $i\omega_n$ for a finite number $N_m$
of consecutive values of the index $n$, i.e., $n=1,2,\ldots,N_m$. This
is, however, not necessary, nor always optimal. In particular, when
the temperature $T$ is very low, the Matsubara points are very dense
and the first $N_m$ points do not necessarily contain enough
information about the high frequency scales, thus it is not possible
to reconstruct the full spectral function. In these circumstances, we
find it convenient to increase the spacing to $s$, so that we use the
imaginary-frequency points 
\begin{equation}
i\omega_n=i(2n+1)s\pi T.
\end{equation}
This corresponds to using the Matsubara points for a higher effective
temperature $s T$, even though the raw spectral data were computed at
the actual physical temperature $T$. The results of this procedure are
illustrated in Fig.~\ref{fig:spectralPade_vs_decimation} for the case
when the temperature is much below the lowest intrinsic temperature
scale of the problem (here the Kondo temperature $T_K=4.6\ 10^{-3}$,
thus $T/T_K=2\ 10^{-4}$). For the standard choice of consecutive
Matsubara points, $s=1$, we find results which are clearly incorrect
at high frequencies, even though the Kondo resonance appears to be
well resolved. For $s=10$, the Hubbard peaks are still not resolved,
but for $s=100$, they are well reproduced. Finally,
for $s=1000$, the spectral function is fully resolved, including the
band-edge features. We note that for $s=1000$ the Matsubara points are
chosen up to the asymptotic $1/z$ tail of the Green's function, see
Fig.\ref{fig:spectralPade_vs_decimation}(c). These results are very
instructive, since they indicate that the information about the
temperature is contained in the raw NRG results, not in the choice of
the Matsubara frequencies. At low physical temperatures it is thus
perfectly safe to use high fictitious temperature in the Pad\'e
approximant calculation.

\begin{figure}[htbp]
\centering
\includegraphics[clip,width=0.48\textwidth]{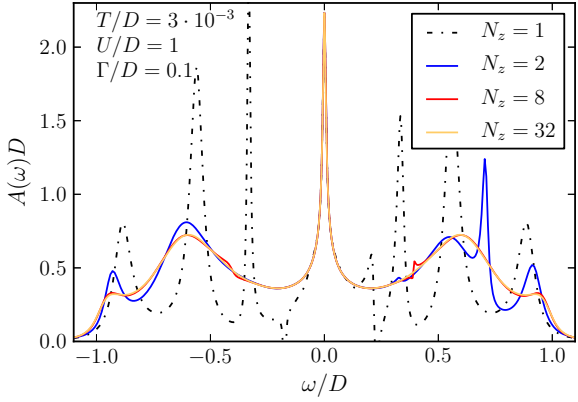}
\caption{(Color online) Spectral function of the SIAM for different
number $N_z$ of the $z$-averaging discretization meshes. For small
$N_z$, the spectra exhibit a large number of spurious resonances
which are eliminated at larger $N_z$.}
\label{fig:Nz}
\end{figure}

It is rather surprising that the Pad\'e approximant produces smooth
spectral functions with relatively little artifacts, given that the
output from the NRG consists of a set of delta peaks at excitations
energies which tend to be clustered. We indeed find that the Pad\'e
fitting applied to the results from a single NRG run with a specific
discretization mesh (i.e., $N_z=1$) produces meaningless results, see
Fig.~\ref{fig:Nz}. However, already at $N_z=2$ the results are
tremendously improved: the Kondo resonance, the Hubbard peaks, and the
band-edge features are all resolved, located at the proper positions
and with roughly correct spectral widths, although there are still
sizable spurious spectral peaks. With finer $z$-averaging, at $N_z=4$
and $N_z=8$, the results are almost converged with only some minor
artifacts. At $N_z=32$ and beyond, we find no further improvement.

\begin{figure}[htbp]
\centering
\includegraphics[clip,width=0.48\textwidth]{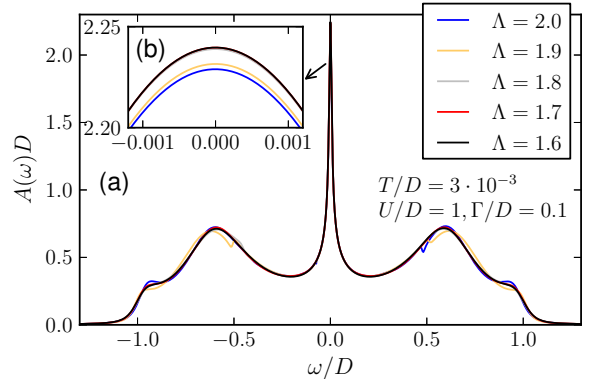}
\caption{(Color online) Spectral function of the SIAM for different
values of the discretization parameters $\Lambda$.}
\label{fig:Lambda}
\end{figure}

In Fig.~\ref{fig:Lambda} we study the convergence with respect to the
NRG discretization parameter $\Lambda$. The continuum limit is
restored for $\Lambda \to 1$, but practical calculations are only
possible for $\Lambda \gtrsim 1.5$. We find that the artifacts are
indeed reduced for smaller $\Lambda$, in particular in the Hubbard
peak region and nead band edges. The form of the Kondo resonance is
well described by all $\Lambda$ in the range considered, but there
appears to be a small yet systematic trend toward higher Kondo
resonance height for smaller $\Lambda$. We have verified that the
Kondo resonance is adequately described even for much higher
$\Lambda=4$, but the description of the Hubbard peaks becomes
increasingly poor (results not shown).

\subsection{The Kondo resonance}
\label{TKsec}

\begin{figure}[htbp]
\centering
\includegraphics[clip,width=0.48\textwidth]{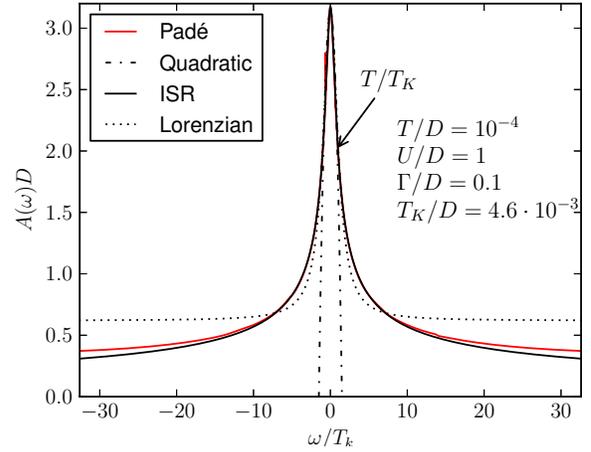}
\caption{(Color online) Shape of the Kondo resonance for $T \ll T_K$: close-up on the
resonance peak and fits to various analytic expressions. 
ISR stands for the inverse-square-root (Frota) form.
}
\label{fig:spectralBroadening_vs_Pade_zoomPeakFit}
\end{figure}

We now discuss the shape of the Kondo resonance, which is well known
to deviate strongly from the Lorenzian form, see
Fig.~\ref{fig:spectralBroadening_vs_Pade_zoomPeakFit}. The behavior
near $\omega=0$ is parabolic, as expected for a regular FL system
\cite{hewson,nozieres1974}. The quadratic fit, however, is only valid
asymptotically for $|\omega| \ll T_K$. The fit to a Lorentzian
function is valid in a somewhat wider energy interval, but it starts
to deviate appreciably already at $\omega \sim T_K$. The long tails
are better approximated by an inverse-square-root function (also known
as the Doniach-\v{S}unji\'c form)
\cite{frota1986,frota1992,bulla2000}, as expected from the
orthogonality-catastrophe physics for the scattering phase-shift
$\pi/2$. The expression due to H. Frota is \cite{frota1992}
\begin{equation}
A(\omega)=\frac{1}{\pi \Gamma} \left(
\frac{1+\sqrt{1+(\omega/\Gamma_K)^2}}{2[1+(\omega/\Gamma_K)^2]}
\right)^{1/2}.
\end{equation}
Using the local-moment approximation method it has been
shown that the tails are asymptotically logarithmic
\cite{logan1998,dickens2001}.

At finite temperatures, the Kondo resonance can again be described by
the Frota form using temperature-dependent parameters. We write
\begin{equation}
A(\omega,T)=\frac{h(T)}{\pi \Gamma} 
\left(
  \frac {1+\sqrt{1+[\omega/\Gamma_K(T)]^2}}
        {2 \left( 1+[\omega/\Gamma_K(T)]^2 \right)}
\right)^{1/2}.
\end{equation}
For the temperature-dependent height we find
\begin{equation}
h(T)=\left[ 1+(2^{1/s}-1)(r T/T_K)^p \right]^{-s},
\end{equation}
with $s=0.29$, $p=1.67$, and $r=0.66$, while for the width we obtain
\begin{equation}
\Gamma_K(T) = c T_K \left(1+ a (T/T_K)^b\right),
\end{equation}
with $c=0.60$, $a=1.97$, and $b=1.33$. This expression well describes
the Kondo resonance in the parameter range $|\omega|<10T_K$ and
$T<10T_K$. Note that we do not fix $p=2$, as would be expected for a
FL system. The $T \to 0$ asymptotic behavior is thus strictly
speaking incorrect. Nevertheless, this choice gives a better
description on the crossover scale $T \gtrsim T_K$, which is more
important for fitting experimental data.

\begin{figure}[htbp]
\centering
\includegraphics[clip,width=0.48\textwidth]{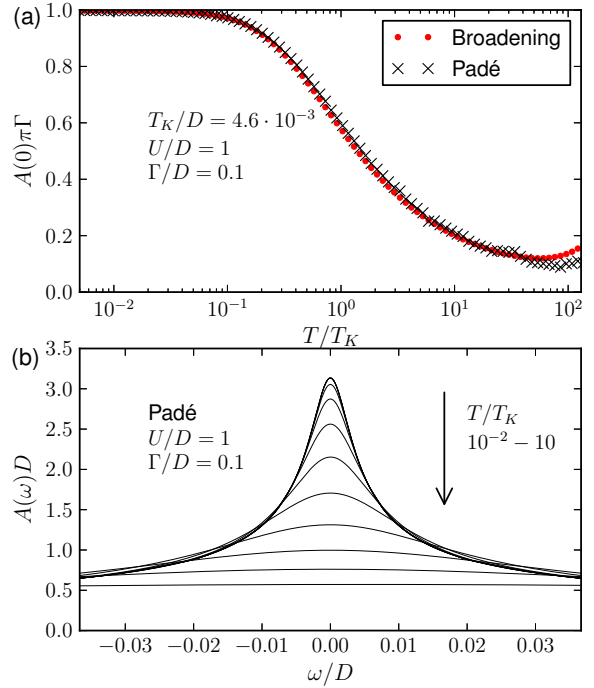}
\caption{(Color online) (a) Kondo resonance height as a function of temperature,
calculated using broadening and Pad\'e approximation.
(b) Spectra for a range of temperatures from
$T/T_K=10^{-2}$ to $T/T_K=10$ in $13$ equally spaced temperatures in the logarithmic scale.
}
\label{fig:peakHeight_vs_T}
\end{figure}

We test how well the Friedel sum rule $\pi\Gamma A(\omega=0)=1$ for
the particle-hole symmetric case is fulfilled in the low-temperature
limit. The results for $\pi\Gamma A(\omega=0,T)$ are shown in
Fig.~\ref{fig:peakHeight_vs_T} on the logarithmic temperature scale.
The sum rule at $T\to 0$ is fulfilled within $0.0018$ using the
broadening approach, and within $0.0019$ using the Pad\'e approximant.
The largest differences between the two approaches are in the
intermediate temperature region where the height of the Kondo
resonance is systematically underestimated by broadening.

Finally, we discuss an integrated spectral quantity, the conductance
through a nanoscopic device (quantum dot) described by the SIAM, as
given by the Meir-Wingreen formula \cite{meir1992}:
\begin{equation}
G(T)=G_0 \int_{-\infty}^{+\infty} \mathrm{d}\omega
\left( -\frac{\partial f}{\partial \omega} \right)
\pi \Gamma A(\omega,T),
\end{equation}
where $G_0=\frac{2e^2}{h}$ is the conductance quantum, and $f$ is the
Fermi-Dirac function. This quantity can be computed directly from raw
spectral data, or using the continuous spectral function obtain either
by broadening or by Pad\'e method. We find that the results agree very
well in all three cases. We fit them on the phenomenological function
proposed by Goldhaber-Gordon et al.
\cite{goldhabergordon1998b,parks2010}:
\begin{equation}
G(T)=G_0 \left[ 1 + (2^{1/s}-1)(T/T_K)^p \right]^{-s}.
\end{equation}
Note that $G(T=T_K)=G_0/2$, which defines the Kondo temperature used
in this work. This definition was used by Hamann, Nagaoka, and Suhl;
it is sometimes denoted as $T_{K,H}$. In addition, there is Wilson's
thermodynamic definition $\chi(T=T_{K,W})/(g\mu_B)^2=0.07$, or the
definition commonly used in the perturbation theory works which is
denoted as $T_K^{(0)}$. They are related through $T_K^{(0)}=0.4128
T_{K,W}$ and $T_{K,H}=2.2 T_{K,W}$. As discussed above, in a FL
system, one may choose to fix $p=2$ in order to obtain the expected $T
\ll T_K$ asymptotics. For broadened spectra we find
\begin{equation}
T_K=4.4\cdot{}10^{-3},\quad s=0.229,
\end{equation}
while for Pad\'e spectra
\begin{equation}
T_K=4.5\cdot{}10^{-3},\quad s=0.227.
\end{equation}
The fitting was performed in the interval from $T/T_K=10^{-2}$ to
$T/T_K=10$. These values agree with the standard result $s=0.23$
\cite{goldhabergordon1998b}. By relaxing the constraint $p=2$ a better
fit is obtained in the intermediate temperature region where the
cross-over occurs. In this case we find for broadened spectra
\begin{equation}
T_K=4.5\cdot{}10^{-3},\quad s=0.277,\quad p=1.74,
\end{equation}
while for Pad\'e spectra
\begin{equation}
T_K=4.6\cdot{}10^{-3},\quad s=0.279,\quad p=1.72.
\end{equation}

\begin{figure}[htbp]
\centering
\includegraphics[clip,width=0.48\textwidth]{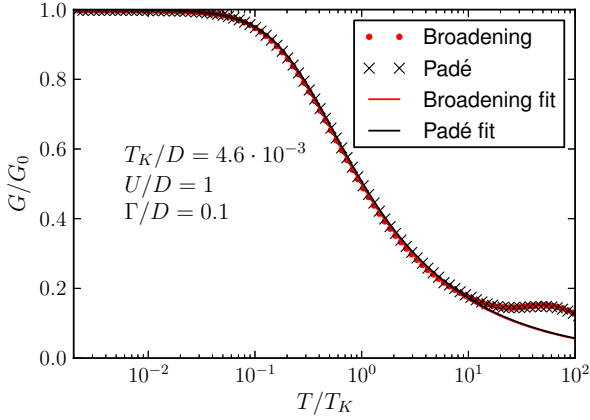}
\caption{(Color online) Conductance $G(T)/G_0$ of a quantum dot
described by the single-impurity Anderson model and a fit to the
phenomenological Goldhaber-Gordon et al. formula.}
\label{fig:conductance_vs_T}
\end{figure}

\subsection{Dynamic susceptibilities}

The spin and charge susceptibilities are defined as
\begin{equation}
\chi_s(z) = \corr{ S_z ; S_z }_z,
\quad
\chi_c(z) = \frac{1}{4} \corr{ n ; n }_z,
\end{equation}
where $S_z=(1/2)(n_\uparrow-n_\downarrow)$ and
$n=n_\uparrow+n_\downarrow$, $n_\sigma$ being the impurity occupancy
operators. We use units such that $g\mu_B=1$, where $g$ is the $g$
ratio and $\mu_B$ is the Bohr magneton, and we set the electron charge
$e=1$. The factor $1/4$ in $\chi_c$ is added for convenience to make
the two susceptibilities equal in the non-interacting limit $U=0$ for
the symmetric case.

For $U=0$, the susceptibilities can be calculated exactly in
terms of pair propagators
\begin{equation}
\begin{split}
\Pi^{p\sigma}_{h\sigma'}(\omega) &= \frac{i}{2\pi}
\int_{-\infty}^{\infty} G_\sigma(\omega+\omega') G_{\sigma'}(\omega')
\mathrm{d}\omega', \\
\Pi^{p\sigma}_{p\sigma'}(\omega) &= \frac{i}{2\pi}
\int_{-\infty}^{\infty} G_\sigma(\omega-\omega') G_{\sigma'}(\omega')
\mathrm{d}\omega',
\end{split}
\end{equation}
as
\begin{equation}
\chi_s(\omega)=\frac{1}{2} \Pi^{p\uparrow}_{h\downarrow}(\omega),
\quad
\chi_c(\omega)=\frac{1}{2} \Pi^{p\uparrow}_{p\downarrow}(\omega).
\end{equation}
The Green's functions here are time ordered, not retarded. The
analytic expression in the wide-band limit are given in
Ref.~\onlinecite{hewson2006spin}. For a finite flat band they need to
be computed numerically. Usually we plot the imaginary parts denoted
as $\chi''(\omega)=\Im[\chi(\omega)]$, i.e., there is no $-1/\pi$
factor as in the spectral function.

The Korringa-Shiba relation is a statement about the spin and charge
dynamical susceptibilities in the low frequency limit. It can be
written as
\begin{equation}
\lim_{\omega \to 0} \frac{\Im \chi(\omega)}{\omega}
= p [\Re\chi(0)]^2,
\label{KS}
\end{equation}
where $p$ is a constant exactly equal to $2\pi$ in the wide-band limit
\cite{shiba1975,hewson2006spin} ($\Gamma,U \ll D$ in SIAM, or $J \ll
D$ in the Kondo model). For a band of finite width, the Korringa-Shiba
relation takes a slightly different form \cite{shiba1975}:
\begin{equation}
\lim_{\omega \to 0} \frac{\Im \chi(\omega)}{\omega} = 2\pi \left[
\Re\corr{S_z ; S_z + S_\mathrm{band} }_{0} \right ]^2,
\label{KS2}
\end{equation}
where $S_\mathrm{band}$ is the spin of the conduction-band electrons,
thus the correlator on the right-hand-side of Eq.~\eqref{KS2} is the
impurity magnetization induced by a small magnetic field applied to
all electrons in the system \cite{shiba1975}. (See
Ref.~\onlinecite{Merker:2012fb} for a related discussion regarding the
thermodynamic susceptibility and the Clogston-Anderson compensation
theorem.) Alternatively, assuming proportionality between the
magnetization in the impurity and in the band, one may work with
Eq.~\eqref{KS} using an effective parameter $p$. For instance, for
non-interacting resonant-level model with $\epsilon=0$ and
$\Gamma/D=10^{-1}$, one finds $p \approx 0.889 \times 2\pi$, while for
$\Gamma/D=10^{-2}$, $p \approx 0.987 \times 2\pi$. The correction to
$p/(2\pi)$ is thus approximately proportional to $\Gamma/D$.

\begin{figure}[htbp]
\centering
\includegraphics[clip,width=0.48\textwidth]{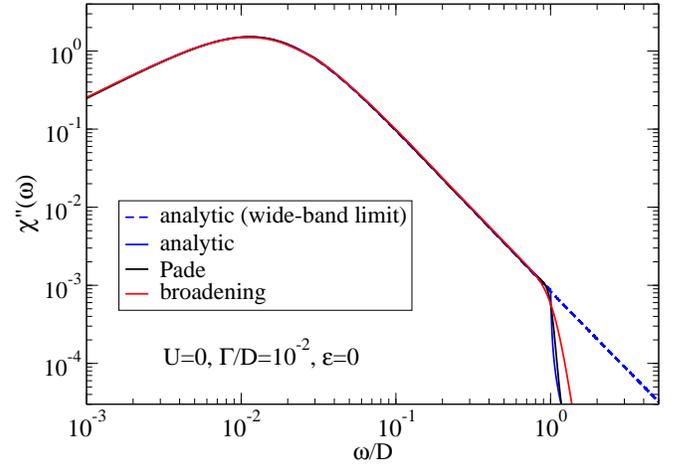}
\caption{(Color online) Dynamical charge susceptibility of the
non-interacting resonant-level model computed with broadening and with
the Pad\'e continuation. We also compare analytic results for the
wide-band limit ($D \to \infty$, dashed line) and finite flat-band
(full line). $T/D=10^{-4}$.}
\label{fig:new1}
\end{figure}

We first test the calculation of susceptibilities on the $U=0$ model,
for a flat band with $\Gamma/D=10^{-2}$. Spin and charge
susceptibilities are the same in this case,
$\chi_s(\omega)=\chi_c(\omega)$. In Fig.~\ref{fig:new1} we compare the
NRG results for $\chi''$ obtained with broadening and Pad\'e, as well
as the analytic results in the wide-band limit and for the actual flat
band. We observe that the susceptibility function has a peak
associated with the charge fluctuation scale of $\Gamma$ and a sharp
drop at the band edge $D$. The analytic result obtained in the
wide-band limit obviously does not account for the drop, which also
explains the difference between the wide-band-limit coefficient $2\pi$
and the effective coefficient $p=0.987 \times 2\pi$. The two numerical
results overlap to a high degree, except near the band edge where the
Pad\'e approximant better describes the sharp decrease. We find that
the NRG result for the slope $\Im\chi(\omega)/\omega$ agrees within
one per mil with the exact result, while $\Re\chi(0)$ obtained via the
Kramers-Kronig transformation of $\chi''$ has a one percent error. We
thus obtain $p=0.967 \times 2\pi$, which is two percent lower than the exact
result. Of course, $\Re\chi(0)$ may be determined more accurately
in NRG by other means.

\begin{figure}[htbp]
\centering
\includegraphics[clip,width=0.48\textwidth]{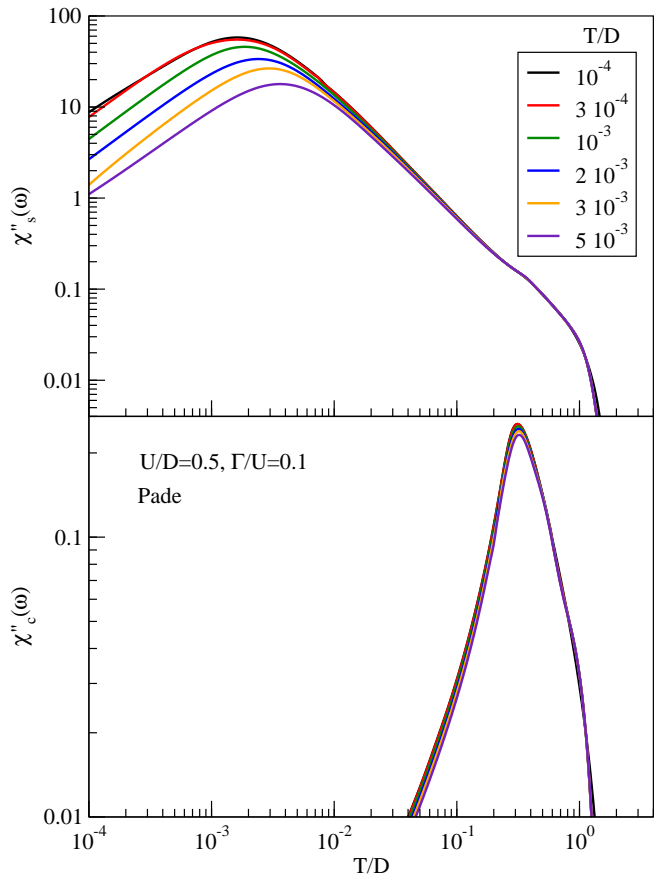}
\caption{(Color online) Dynamical spin (top) and charge (bottom)
susceptibility of the single-impurity Anderson model obtained with
Pad\'e continuation for a range of temperatures. The Kondo temperature
is $T_K=2.5\ 10^{-3}D$.}
\label{fig:news}
\end{figure}

In Fig.~\ref{fig:news} we plot the dynamical spin and charge
susceptibilities for the interacting SIAM in the Kondo regime for a
range of temperatures. The charge susceptibility has a dominant peak
on the scale of $U$ (the maximum is at $\omega \approx 0.64 U$), and a
slight change of slope at the band edge $\omega=D$. The spin
susceptibility has a dominant peak associated with spin fluctuations
on the scale of $T_K$, a hint of a spectral feature at the frequency
where the charge susceptibility has a maximum, and a drop at the band
edge $D$. With increasing temperature (in the range $T\sim T_K$), the
charge susceptibility peak diminishes slightly, while the main peak in
the spin susceptibility shifts to higher frequencies and decreases in
magnitude. We observe that the Pad\'e approximant approach has
difficulties with producing correct slopes in the $\omega \to 0$ limit
[see, for example, the $T/D=3\ 10^{-3}$ curve in
Fig.~\ref{fig:news}(a)]. In order to determine the true origin of the
problem, we studied the raw binned data and plotted the cumulant
function 
\begin{equation}
\int_0^x \chi''_\mathrm{NRG}(\omega) \mathrm{d}\omega,
\end{equation}
which is relatively smooth for large $N_z$. If $\chi''(\omega)$ truly
behaved linearly near $\omega=0$, the cumulant should be quadratic for
small $x$. We confirmed that this holds to a good approximation for
$x$ somewhat lower than the lowest energy scale of the problem
($\Gamma$ for the non-interacting case, $T_K$ for the SIAM), but not
for $x$ lower than the temperature $T$, where we find different power
laws and the slope extraction becomes unreliable. This indicates that
the Korringa-Shiba relation can only be tested in the $T \ll \Gamma$
and $T \ll T_K$ limits, respectively, but not for high temperatures
where there is too much uncertainty. It also indicates that the raw
dynamic information in the FDM NRG approach for $\omega < T$ is not
reliable. While this does not appear to be an issue for spectral
functions, where the Pad\'e approach produces what seems to be a
reliable fit to the available raw data, this is not the case for
dynamical susceptibilities where we observe incorrect slopes at low
frequencies.

\section{Results for a  correlated-electrons problem: Hubbard model
within the DMFT}
\label{secdmft}

We apply the Pad\'e approximant approach to determine the spectral
function of the Hubbard model in the paramagnetic phase within the
DMFT. The questions of main interest are: i) Can we extend the upper
limit of the temperature range where the DMFT(NRG) method can be
safely applied towards $T\sim D$ or even beyond?  ii) Can we obtain
more detailed information about the internal structure of the Hubbard
bands? iii) Is the problem of
causality violation at lower temperatures reduced?

We use the Bethe lattice with the non-interacting Green's function
\begin{equation}
G_0(z)=\frac{2}{D} \left[ \frac{z}{D}-i \sgn \left(\Im{z} \right) \sqrt{1-{\frac{z^2}{D^2}}} \right].
\end{equation}
The DOS of the Bethe lattice shares some features with the 3D cubic
lattice DOS. For instance, at the band edges it has square root
singularities. The calculations are performed with $N_z=8$ and we take
advantage of the Broyden mixing to improve the convergence. 

\begin{figure}[htbp]
\centering
\includegraphics[clip,width=0.48\textwidth]{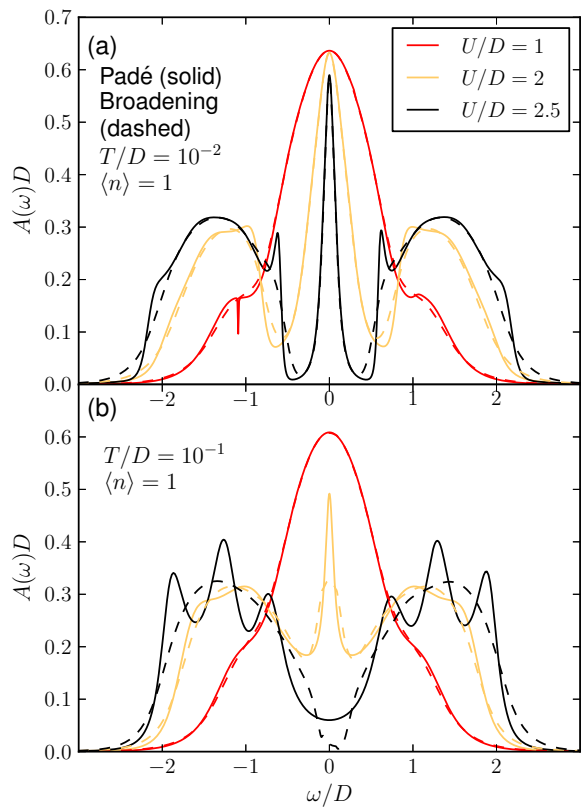}
\caption{(Color online) Spectral functions of the particle-hole
symmetric Hubbard model at half-filling, $\langle{}n\rangle=1$, at
temperatures (a) $T/D=0.01$ and (b) $T/D=0.1$ for a range of repulsion
strengths $U/D$ computed using the dynamical mean-field theory (DMFT).
We compare spectra computed using the Pad\'e approach (solid lines)
and with broadening (dashed lines). (Pad\'e continuation is used in
all steps of the DMFT calculation, not just to obtain the final
spectrum.)
}
\label{fig:betheSpectral_n1_vs_U}
\end{figure}

We first discuss the half-filled system, $\expv{n}=1$, which is
particle-hole symmetric. The local spectral functions are shown in
Fig.~\ref{fig:betheSpectral_n1_vs_U} for two different temperatures.
For low and moderate interaction $U$, the behavior is qualitatively
the same as in the non-self-consistent SIAM. The Kondo resonance at
low frequencies is reinterpreted as the quasiparticle band and there
are two Hubbard bands which now exhibit some internal structure, in
particular some enhancement at the inner band edges, which has also
been resolved previously in the broadening approach with very narrow
kernels \cite{resolution} and is known to be a real feature of the DOS
\cite{karski2005,karski2008}; see, for example, the $U/D=2.5$ curve at
$T/D=10^{-2}$ in Fig.~\ref{fig:betheSpectral_n1_vs_U}.

The spectra at higher temperature, $T/D=10^{-1}$, are shown in the
lower panel of Fig.~\ref{fig:betheSpectral_n1_vs_U}. For values of $U$
approaching the $T=0$ Mott metal-insulator transition, the
quasiparticle peak is strongly suppressed and the spectral
distribution inside the Hubbard bands is modified (see $U/D=2$
spectra). Broadening seems to underestimate the height of the
quasiparticle peak in this regime. For larger $U/D=2.5$, the
quasiparticle peak is completely washed out and the Pad\'e results
suggest that the Hubbard bands develop a pronounced three peak
structure. This is an artifact of the method and broadening results
show no such structure.

\begin{figure}[htbp]
\centering
\includegraphics[clip,width=0.48\textwidth]{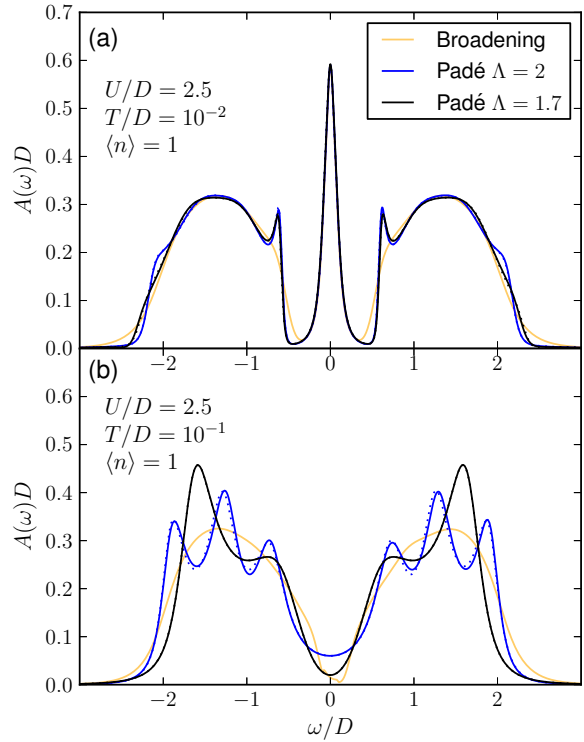}
\caption{(Color online) Spectral functions of the particle-hole
symmetric Hubbard model at half-filling for $U/D=2.5$: dependence on
the NRG discretization parameter $\Lambda$. Dashed line show $A(-\omega)D$.
}
\label{fig:U25}
\end{figure}

In Fig.~\ref{fig:U25} we explore more carefully the reliability of the
Pad\'e approach in determining the internal structure of the Hubbard
bands. In (a), we show a low temperature result where the Pad\'e
approach at $\Lambda=2$ suggests, in addition to the well defined
resonance at the inner band edge, also some feature at the outer
Hubbard band edge. By reducing the discretization to $\Lambda=1.7$,
this feature is washed out, which indicates that it was an NRG
artifact. This is a general recepe: real spectral features may be
distinguished from artifacts by changing the NRG parameters (such as
$\Lambda$, truncation cutoffs, etc.), whereby real features should
remain robust, while artifacts are very variable. This is well
illustrated in (b), where the results confirm that the three peek
structure observed for $U/D=2.5$ at higher temperature is indeed an
artifact.

\begin{figure}[htbp]
\centering
\includegraphics[clip,width=0.48\textwidth]{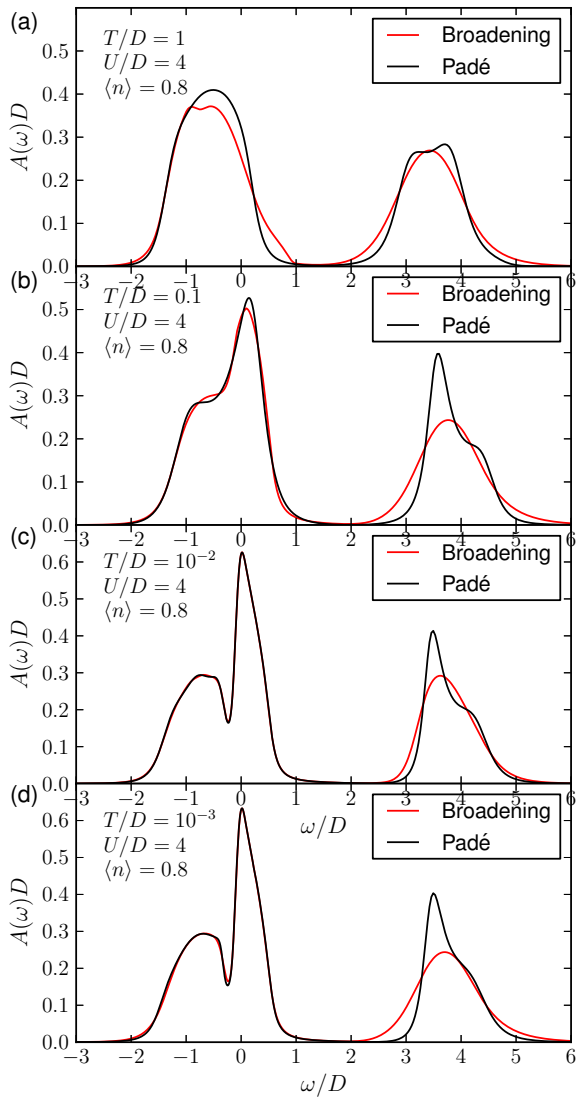}
\caption{(Color online) Spectral functions of the Hubbard model
describing a doped Mott insulator for a range of temperatures. We show
curves obtained by broadening ($\alpha=0.2$) and Pad\'e approximation.
}
\label{fig:betheSpectral_vs_Method}
\end{figure}

We next consider a hole-dopped Mott insulator with $\expv{n}=0.8$ at
$U/D=4$. This same parameter set has also been used in a recent study
of the transport properties of this model, see Ref.~\onlinecite{deng}.
For low temperatures we find very good agreement between broadening
and Pad\'e approaches for the low-frequency part of the spectra, while
at high frequencies the Pad\'e approach reveals some internal
structure of the upper Hubbard band which can only be obtained in the
broadening approach if the broadening parameter $\alpha$ is much
reduced. We note, however, that the integrated difference between two
sequential Green's function in the DMFT loop does not go to zero in
the Pad\'e approach but rather oscillates around $10^{-6}$. This lack
of true convergence can be shown to originate precisely in the inner
structure of the upper Hubbard peak, which changes subtly from
iteration to iteration. Such behavior is sometimes associated with
physical instabilities of the system which are not allowed for in the
DMFT Ansatz, but may also here be an artifact. In any case, we believe
that the upper Hubbard band does have internal structure and differs
from simple non-interacting DOS.

We finally address the problems in the DMFT(NRG) at low temperatures
which manifest as the violation of the causality in the calculated
self-energy functions, as discussed in the introduction. The self-energy
has different asymptotic behavior than the hybridization function or 
the Green's function, so special care must be taken when performing
continuation with Pad\'e. For large frequencies, it behaves as
\cite{gull2011}
\begin{equation}
\Sigma(z) \sim U\langle{}n\rangle{}/2 + O(1/z).
\end{equation}
Because Pad\'e requires $O(1/z)$ asymptotic behavior, we first subtract 
$U\langle{}n\rangle{}/2$, perform the continuation and add the 
subtracted value back to the result. 
It was believed that since the self-energy is a ratio
of two Green's function and the real parts need to be obtained
by the Kramers-Kronig procedure, it was supposed that the
(over)broadening of the spectral functions (imaginary parts) at high
frequencies has an impact on all frequency scales in the real parts,
thus spoiling the causality. This appears not to be the case, since in
the Pad\'e approximant approach the problems begin to show at the same
temperature scale as with broadening. 
The causality violation thus appears to be intrinsic to the
self-energy trick in NRG.

Interestingly, we did not observe similar problems with the
self-energy in calculations for the SIAM, but only in self-consistent
DMFT calculations.

\section{Conclusion}

We have analyzed a non-standard approach for calculating
finite-temperature spectral functions using the full-density-matrix 
numerical renormalization group (FDM NRG). It consists of evaluating
the Green's function on the imaginary frequency axis, either at the
Matsubara frequencies $i\omega_n=i(2n+1) \pi T$ or at some other
conveniently chosen set of points. To obtain the spectral function on
the real frequency axis, an analytic continuation using the Pad\'e
approximant is performed. The technique works well in conjunction with
sufficient $z$-averaging, which partially removes the discretization
artifacts of the NRG. Compared with the broadening approach, the
technique does not have any arbitrariness in the choice of the
broadening kernel and the only free parameter in most circumstances is
the number of Matsubara points $N_m$; the low-energy part of the
spectral function converges quickly with increasing $N_m$. In
addition, unlike with broadening, the spectral function does not have
artifacts on the frequency scale of $\omega \sim T$, which is
particularly important for calculating transport properties
(conductance, thermopower) of various systems described by impurity
problems, both in the context of nanodevices and in the dynamical
mean-field theory (DMFT). Some artifacts in spectral functions appear
on high frequency scales and we have shown that their origin is in the
raw NRG data, not in the analytic continuation procedure as such. A
further advantage of the proposed technique is its capability to
resolve narrow spectral features away from the Fermi level, unlike in
the broadening approach where the kernel width is usually proportional
to the frequency and may thus far exceed the width of the spectral
feature under consideration. The reduction of overbroadening problems
is sufficiently good that the self-energy trick of calculating the
interaction self-energy function as a ratio of two correlators is not
necessary. We have also shown that the technique may be applied at
relatively low temperatures by choosing instead of the Matsubara
points a modified set of points on the imaginary axis which provides a
good sampling of the spectral information. We have tested the method
first on the single impurity Anderson model, where we have analyzed
the properties of the Kondo resonance both at low and at intermediate
($T \sim T_K$) temperatures. Since the method does not suffer from
overbroadening nor from a lack of smoothness at finite temperatures,
we were able to obtain the results for the Kondo resonance with
unprecedented reliability. We discussed various analytic expression
for the Kondo peak shape and proposed an expression which is valid
in a wide temperature range. We have also tested the approach on the
Hubbard model within the DMFT. We find that the Pad\'e approximant
approach has a twofold advantage over the broadening: i) It has less
spectral-function artifacts on the important scale of $\omega\sim T$.
ii) It can better resolve the inner structure of the Hubbard bands.

The main drawback of the Pad\'e approximant are the artifacts which,
however, can be systematically eliminated by reducing $\Lambda$ and
increasing $N_z$. In addition, we find that while the Pad\'e approach
works well for spectral functions, it has more difficulties with
reproducing correct slopes of dynamical susceptibilities at finite
temperatures (which is an issue of NRG itself).

The importance of this work is not only in proposing an improved
technique for obtaining finite-temperature spectral functions using
Pad\'e approximants, but more generally in showing that broadening of
delta peaks is not necessarily the only nor the best approach. A
possible improvement of our approach would be to build in as
constraints additional information about the spectral function, such
as the spectral moments which are known to very high accuracy, and
obtain the Pad\'e approximant using an optimization procedure. 
One could also consider different fit functions. Work along these
lines is in progress.

\begin{acknowledgments}
We acknowledge discussions with Jernej Mravlje, Michele Fabrizio and
Thomas Pruschke and the support of the Slovenian Research Agency
(ARRS) under Program P1-0044.
\end{acknowledgments}

\bibliography{mats}

\begin{thebibliography}{116}
\expandafter\ifx\csname natexlab\endcsname\relax\def\natexlab#1{#1}\fi
\expandafter\ifx\csname bibnamefont\endcsname\relax
  \def\bibnamefont#1{#1}\fi
\expandafter\ifx\csname bibfnamefont\endcsname\relax
  \def\bibfnamefont#1{#1}\fi
\expandafter\ifx\csname citenamefont\endcsname\relax
  \def\citenamefont#1{#1}\fi
\expandafter\ifx\csname url\endcsname\relax
  \def\url#1{\texttt{#1}}\fi
\expandafter\ifx\csname urlprefix\endcsname\relax\def\urlprefix{URL }\fi
\providecommand{\bibinfo}[2]{#2}
\providecommand{\eprint}[2][]{\url{#2}}

\bibitem[{\citenamefont{Hewson}(1993{\natexlab{a}})}]{hewson}
\bibinfo{author}{\bibfnamefont{A.~C.} \bibnamefont{Hewson}},
  \emph{\bibinfo{title}{The Kondo Problem to Heavy-Fermions}}
  (\bibinfo{publisher}{Cambridge University Press, Cambridge},
  \bibinfo{year}{1993}{\natexlab{a}}).

\bibitem[{\citenamefont{Kondo}(1964)}]{kondo1964}
\bibinfo{author}{\bibfnamefont{J.}~\bibnamefont{Kondo}},
  \bibinfo{journal}{Prog. Theor. Phys.} \textbf{\bibinfo{volume}{32}},
  \bibinfo{pages}{37} (\bibinfo{year}{1964}).

\bibitem[{\citenamefont{Kouwenhoven and Glazman}(2001)}]{kouwenhoven2001}
\bibinfo{author}{\bibfnamefont{L.}~\bibnamefont{Kouwenhoven}} \bibnamefont{and}
  \bibinfo{author}{\bibfnamefont{L.}~\bibnamefont{Glazman}},
  \bibinfo{journal}{Physics World} \textbf{\bibinfo{volume}{14}},
  \bibinfo{pages}{33} (\bibinfo{year}{2001}).

\bibitem[{\citenamefont{Anderson and Yuval}(1969)}]{anderson1969exact1}
\bibinfo{author}{\bibfnamefont{P.~W.} \bibnamefont{Anderson}} \bibnamefont{and}
  \bibinfo{author}{\bibfnamefont{G.}~\bibnamefont{Yuval}},
  \bibinfo{journal}{Phys. Rev. Lett.} \textbf{\bibinfo{volume}{23}},
  \bibinfo{pages}{89} (\bibinfo{year}{1969}).

\bibitem[{\citenamefont{Anderson et~al.}(1970)\citenamefont{Anderson, Yuval,
  and Hamann}}]{anderson1970exact2}
\bibinfo{author}{\bibfnamefont{P.~W.} \bibnamefont{Anderson}},
  \bibinfo{author}{\bibfnamefont{G.}~\bibnamefont{Yuval}}, \bibnamefont{and}
  \bibinfo{author}{\bibfnamefont{D.~R.} \bibnamefont{Hamann}},
  \bibinfo{journal}{Phys. Rev. B} \textbf{\bibinfo{volume}{1}},
  \bibinfo{pages}{4464} (\bibinfo{year}{1970}).

\bibitem[{\citenamefont{Anderson}(1970)}]{anderson1970}
\bibinfo{author}{\bibfnamefont{P.~W.} \bibnamefont{Anderson}},
  \bibinfo{journal}{J. Phys. C: Solid St. Phys.} \textbf{\bibinfo{volume}{3}},
  \bibinfo{pages}{2436} (\bibinfo{year}{1970}).

\bibitem[{\citenamefont{Haldane}(1978)}]{haldane1978}
\bibinfo{author}{\bibfnamefont{F.~D.~M.} \bibnamefont{Haldane}},
  \bibinfo{journal}{Phys. Rev. Lett.} \textbf{\bibinfo{volume}{40}},
  \bibinfo{pages}{416} (\bibinfo{year}{1978}).

\bibitem[{\citenamefont{Nozi{\`e}res}(1974)}]{nozieres1974}
\bibinfo{author}{\bibfnamefont{P.}~\bibnamefont{Nozi{\`e}res}},
  \bibinfo{journal}{J. Low. Temp. Phys.} \textbf{\bibinfo{volume}{17}},
  \bibinfo{pages}{31} (\bibinfo{year}{1974}).

\bibitem[{\citenamefont{Hewson}(1993{\natexlab{b}})}]{hewson1993}
\bibinfo{author}{\bibfnamefont{A.~C.} \bibnamefont{Hewson}},
  \bibinfo{journal}{Phys. Rev. Lett.} \textbf{\bibinfo{volume}{70}},
  \bibinfo{pages}{4007} (\bibinfo{year}{1993}{\natexlab{b}}).

\bibitem[{\citenamefont{Wilson}(1975)}]{wilson1975}
\bibinfo{author}{\bibfnamefont{K.~G.} \bibnamefont{Wilson}},
  \bibinfo{journal}{Rev. Mod. Phys.} \textbf{\bibinfo{volume}{47}},
  \bibinfo{pages}{773} (\bibinfo{year}{1975}).

\bibitem[{\citenamefont{Krishna-murthy
  et~al.}(1975)\citenamefont{Krishna-murthy, Wilkins, and
  Wilson}}]{krishna1975}
\bibinfo{author}{\bibfnamefont{H.~R.} \bibnamefont{Krishna-murthy}},
  \bibinfo{author}{\bibfnamefont{J.~W.} \bibnamefont{Wilkins}},
  \bibnamefont{and} \bibinfo{author}{\bibfnamefont{K.~G.}
  \bibnamefont{Wilson}}, \bibinfo{journal}{Phys. Rev. Lett.}
  \textbf{\bibinfo{volume}{35}}, \bibinfo{pages}{1101} (\bibinfo{year}{1975}).

\bibitem[{\citenamefont{Krishna-murthy
  et~al.}(1980{\natexlab{a}})\citenamefont{Krishna-murthy, Wilkins, and
  Wilson}}]{krishna1980a}
\bibinfo{author}{\bibfnamefont{H.~R.} \bibnamefont{Krishna-murthy}},
  \bibinfo{author}{\bibfnamefont{J.~W.} \bibnamefont{Wilkins}},
  \bibnamefont{and} \bibinfo{author}{\bibfnamefont{K.~G.}
  \bibnamefont{Wilson}}, \bibinfo{journal}{Phys. Rev. B}
  \textbf{\bibinfo{volume}{21}}, \bibinfo{pages}{1003}
  (\bibinfo{year}{1980}{\natexlab{a}}).

\bibitem[{\citenamefont{Krishna-murthy
  et~al.}(1980{\natexlab{b}})\citenamefont{Krishna-murthy, Wilkins, and
  Wilson}}]{krishna1980b}
\bibinfo{author}{\bibfnamefont{H.~R.} \bibnamefont{Krishna-murthy}},
  \bibinfo{author}{\bibfnamefont{J.~W.} \bibnamefont{Wilkins}},
  \bibnamefont{and} \bibinfo{author}{\bibfnamefont{K.~G.}
  \bibnamefont{Wilson}}, \bibinfo{journal}{Phys. Rev. B}
  \textbf{\bibinfo{volume}{21}}, \bibinfo{pages}{1044}
  (\bibinfo{year}{1980}{\natexlab{b}}).

\bibitem[{\citenamefont{Bulla et~al.}(2008)\citenamefont{Bulla, Costi, and
  Pruschke}}]{bulla2008}
\bibinfo{author}{\bibfnamefont{R.}~\bibnamefont{Bulla}},
  \bibinfo{author}{\bibfnamefont{T.}~\bibnamefont{Costi}}, \bibnamefont{and}
  \bibinfo{author}{\bibfnamefont{T.}~\bibnamefont{Pruschke}},
  \bibinfo{journal}{Rev. Mod. Phys.} \textbf{\bibinfo{volume}{80}},
  \bibinfo{pages}{395} (\bibinfo{year}{2008}).

\bibitem[{\citenamefont{Andrei et~al.}(1983)\citenamefont{Andrei, Furuya, and
  Lowenstein}}]{andrei1983}
\bibinfo{author}{\bibfnamefont{N.}~\bibnamefont{Andrei}},
  \bibinfo{author}{\bibfnamefont{K.}~\bibnamefont{Furuya}}, \bibnamefont{and}
  \bibinfo{author}{\bibfnamefont{J.~H.} \bibnamefont{Lowenstein}},
  \bibinfo{journal}{Rev. Mod. Phys.} \textbf{\bibinfo{volume}{55}},
  \bibinfo{pages}{331} (\bibinfo{year}{1983}).

\bibitem[{\citenamefont{Tsvelick and
  Wiegmann}(1983{\natexlab{a}})}]{tsvelick1983}
\bibinfo{author}{\bibfnamefont{A.~M.} \bibnamefont{Tsvelick}} \bibnamefont{and}
  \bibinfo{author}{\bibfnamefont{P.~B.} \bibnamefont{Wiegmann}},
  \bibinfo{journal}{Adv. Phys.} \textbf{\bibinfo{volume}{32}},
  \bibinfo{pages}{453} (\bibinfo{year}{1983}{\natexlab{a}}).

\bibitem[{\citenamefont{Tsvelick and
  Wiegmann}(1983{\natexlab{b}})}]{tsvelick1983td}
\bibinfo{author}{\bibfnamefont{A.~M.} \bibnamefont{Tsvelick}} \bibnamefont{and}
  \bibinfo{author}{\bibfnamefont{P.~B.} \bibnamefont{Wiegmann}},
  \bibinfo{journal}{J. Phys. C: Solid State Phys.}
  \textbf{\bibinfo{volume}{16}}, \bibinfo{pages}{2321}
  (\bibinfo{year}{1983}{\natexlab{b}}).

\bibitem[{\citenamefont{Oliveira and
  Wilkins}(1981{\natexlab{a}})}]{oliveira1981}
\bibinfo{author}{\bibfnamefont{L.~N.} \bibnamefont{Oliveira}} \bibnamefont{and}
  \bibinfo{author}{\bibfnamefont{J.~W.} \bibnamefont{Wilkins}},
  \bibinfo{journal}{Phys. Rev. Lett.} \textbf{\bibinfo{volume}{47}},
  \bibinfo{pages}{1553} (\bibinfo{year}{1981}{\natexlab{a}}).

\bibitem[{\citenamefont{Oliveira and Oliveira}(1994)}]{oliveira1994}
\bibinfo{author}{\bibfnamefont{W.~C.} \bibnamefont{Oliveira}} \bibnamefont{and}
  \bibinfo{author}{\bibfnamefont{L.~N.} \bibnamefont{Oliveira}},
  \bibinfo{journal}{Phys. Rev. B} \textbf{\bibinfo{volume}{49}},
  \bibinfo{pages}{11986} (\bibinfo{year}{1994}).

\bibitem[{\citenamefont{Oliveira and
  Wilkins}(1981{\natexlab{b}})}]{oliveira1981phaseshift}
\bibinfo{author}{\bibfnamefont{L.~N.} \bibnamefont{Oliveira}} \bibnamefont{and}
  \bibinfo{author}{\bibfnamefont{J.~W.} \bibnamefont{Wilkins}},
  \bibinfo{journal}{Phys. Rev. B} \textbf{\bibinfo{volume}{24}},
  \bibinfo{pages}{4863} (\bibinfo{year}{1981}{\natexlab{b}}).

\bibitem[{\citenamefont{Frota and Oliveira}(1986)}]{frota1986}
\bibinfo{author}{\bibfnamefont{H.~O.} \bibnamefont{Frota}} \bibnamefont{and}
  \bibinfo{author}{\bibfnamefont{L.~N.} \bibnamefont{Oliveira}},
  \bibinfo{journal}{Phys. Rev. B} \textbf{\bibinfo{volume}{33}},
  \bibinfo{pages}{7871} (\bibinfo{year}{1986}).

\bibitem[{\citenamefont{Sakai et~al.}(1989)\citenamefont{Sakai, Shimizu, and
  Kasuya}}]{sakai1989}
\bibinfo{author}{\bibfnamefont{O.}~\bibnamefont{Sakai}},
  \bibinfo{author}{\bibfnamefont{Y.}~\bibnamefont{Shimizu}}, \bibnamefont{and}
  \bibinfo{author}{\bibfnamefont{T.}~\bibnamefont{Kasuya}},
  \bibinfo{journal}{J. Phys. Soc. Jpn.} \textbf{\bibinfo{volume}{58}},
  \bibinfo{pages}{3666} (\bibinfo{year}{1989}).

\bibitem[{\citenamefont{Yoshida et~al.}(1990)\citenamefont{Yoshida, Whitaker,
  and Oliveira}}]{yoshida1990}
\bibinfo{author}{\bibfnamefont{M.}~\bibnamefont{Yoshida}},
  \bibinfo{author}{\bibfnamefont{M.~A.} \bibnamefont{Whitaker}},
  \bibnamefont{and} \bibinfo{author}{\bibfnamefont{L.~N.}
  \bibnamefont{Oliveira}}, \bibinfo{journal}{Phys. Rev. B}
  \textbf{\bibinfo{volume}{41}}, \bibinfo{pages}{9403} (\bibinfo{year}{1990}).

\bibitem[{\citenamefont{Costi and Hewson}(1992)}]{costi1991}
\bibinfo{author}{\bibfnamefont{T.~A.} \bibnamefont{Costi}} \bibnamefont{and}
  \bibinfo{author}{\bibfnamefont{A.~C.} \bibnamefont{Hewson}},
  \bibinfo{journal}{Phil. Mag. B} \textbf{\bibinfo{volume}{65}},
  \bibinfo{pages}{1165} (\bibinfo{year}{1992}).

\bibitem[{\citenamefont{Sakai et~al.}(1992)\citenamefont{Sakai, Shimizu, and
  Kasuya}}]{sakai1992siam}
\bibinfo{author}{\bibfnamefont{O.}~\bibnamefont{Sakai}},
  \bibinfo{author}{\bibfnamefont{Y.}~\bibnamefont{Shimizu}}, \bibnamefont{and}
  \bibinfo{author}{\bibfnamefont{T.}~\bibnamefont{Kasuya}},
  \bibinfo{journal}{Prog. theor. phys.} \textbf{\bibinfo{volume}{108}},
  \bibinfo{pages}{73} (\bibinfo{year}{1992}).

\bibitem[{\citenamefont{Costi and Hewson}(1993)}]{costi1993}
\bibinfo{author}{\bibfnamefont{T.~A.} \bibnamefont{Costi}} \bibnamefont{and}
  \bibinfo{author}{\bibfnamefont{A.~C.} \bibnamefont{Hewson}},
  \bibinfo{journal}{J. Phys. - Cond. Mat.} \textbf{\bibinfo{volume}{5}},
  \bibinfo{pages}{L361} (\bibinfo{year}{1993}).

\bibitem[{\citenamefont{Costi et~al.}(1994)\citenamefont{Costi, Hewson, and
  Zlati{\'c}}}]{costi1994}
\bibinfo{author}{\bibfnamefont{T.~A.} \bibnamefont{Costi}},
  \bibinfo{author}{\bibfnamefont{A.~C.} \bibnamefont{Hewson}},
  \bibnamefont{and}
  \bibinfo{author}{\bibfnamefont{V.}~\bibnamefont{Zlati{\'c}}},
  \bibinfo{journal}{J. Phys.: Condens. Matter} \textbf{\bibinfo{volume}{6}},
  \bibinfo{pages}{2519} (\bibinfo{year}{1994}).

\bibitem[{\citenamefont{Andergassen et~al.}(2011)\citenamefont{Andergassen,
  Costi, and Zlati\'c}}]{andergassen2011}
\bibinfo{author}{\bibfnamefont{S.}~\bibnamefont{Andergassen}},
  \bibinfo{author}{\bibfnamefont{T.~A.} \bibnamefont{Costi}}, \bibnamefont{and}
  \bibinfo{author}{\bibfnamefont{V.}~\bibnamefont{Zlati\'c}},
  \bibinfo{journal}{Phys. Rev. B} \textbf{\bibinfo{volume}{84}},
  \bibinfo{pages}{241107} (\bibinfo{year}{2011}).

\bibitem[{\citenamefont{Rejec et~al.}(2012)\citenamefont{Rejec, \v{Z}itko,
  Mravlje, and Ram\v{s}ak}}]{rejec2012}
\bibinfo{author}{\bibfnamefont{T.}~\bibnamefont{Rejec}},
  \bibinfo{author}{\bibfnamefont{R.}~\bibnamefont{\v{Z}itko}},
  \bibinfo{author}{\bibfnamefont{J.}~\bibnamefont{Mravlje}}, \bibnamefont{and}
  \bibinfo{author}{\bibfnamefont{A.}~\bibnamefont{Ram\v{s}ak}},
  \bibinfo{journal}{Phys. Rev. B} \textbf{\bibinfo{volume}{85}},
  \bibinfo{pages}{085117} (\bibinfo{year}{2012}).

\bibitem[{\citenamefont{Bulla et~al.}(1998)\citenamefont{Bulla, Hewson, and
  Pruschke}}]{bulla1998}
\bibinfo{author}{\bibfnamefont{R.}~\bibnamefont{Bulla}},
  \bibinfo{author}{\bibfnamefont{A.~C.} \bibnamefont{Hewson}},
  \bibnamefont{and} \bibinfo{author}{\bibfnamefont{T.}~\bibnamefont{Pruschke}},
  \bibinfo{journal}{J. Phys.: Condens. Matter} \textbf{\bibinfo{volume}{10}},
  \bibinfo{pages}{8365} (\bibinfo{year}{1998}).

\bibitem[{\citenamefont{Hofstetter}(2000)}]{hofstetter2000}
\bibinfo{author}{\bibfnamefont{W.}~\bibnamefont{Hofstetter}},
  \bibinfo{journal}{Phys. Rev. Lett.} \textbf{\bibinfo{volume}{85}},
  \bibinfo{pages}{1508} (\bibinfo{year}{2000}).

\bibitem[{\citenamefont{Campo and Oliveira}(2005)}]{campo2005}
\bibinfo{author}{\bibfnamefont{V.~L.} \bibnamefont{Campo}} \bibnamefont{and}
  \bibinfo{author}{\bibfnamefont{L.~N.} \bibnamefont{Oliveira}},
  \bibinfo{journal}{Phys. Rev. B} \textbf{\bibinfo{volume}{72}},
  \bibinfo{pages}{104432} (\bibinfo{year}{2005}).

\bibitem[{\citenamefont{Anders and Schiller}(2005)}]{anders2005}
\bibinfo{author}{\bibfnamefont{F.~B.} \bibnamefont{Anders}} \bibnamefont{and}
  \bibinfo{author}{\bibfnamefont{A.}~\bibnamefont{Schiller}},
  \bibinfo{journal}{Phys. Rev. Lett.} \textbf{\bibinfo{volume}{95}},
  \bibinfo{pages}{196801} (\bibinfo{year}{2005}).

\bibitem[{\citenamefont{Anders and Schiller}(2006)}]{anders2006}
\bibinfo{author}{\bibfnamefont{F.~B.} \bibnamefont{Anders}} \bibnamefont{and}
  \bibinfo{author}{\bibfnamefont{A.}~\bibnamefont{Schiller}},
  \bibinfo{journal}{Phys. Rev. B} \textbf{\bibinfo{volume}{74}},
  \bibinfo{pages}{245113} (\bibinfo{year}{2006}).

\bibitem[{\citenamefont{Peters et~al.}(2006)\citenamefont{Peters, Pruschke, and
  Anders}}]{peters2006}
\bibinfo{author}{\bibfnamefont{R.}~\bibnamefont{Peters}},
  \bibinfo{author}{\bibfnamefont{T.}~\bibnamefont{Pruschke}}, \bibnamefont{and}
  \bibinfo{author}{\bibfnamefont{F.~B.} \bibnamefont{Anders}},
  \bibinfo{journal}{Phys. Rev. B} \textbf{\bibinfo{volume}{74}},
  \bibinfo{pages}{245114} (\bibinfo{year}{2006}).

\bibitem[{\citenamefont{\v{Z}itko and Pruschke}(2009)}]{resolution}
\bibinfo{author}{\bibfnamefont{R.}~\bibnamefont{\v{Z}itko}} \bibnamefont{and}
  \bibinfo{author}{\bibfnamefont{T.}~\bibnamefont{Pruschke}},
  \bibinfo{journal}{Phys. Rev. B} \textbf{\bibinfo{volume}{79}},
  \bibinfo{pages}{085106} (\bibinfo{year}{2009}).

\bibitem[{\citenamefont{Weichselbaum and von Delft}(2007)}]{weichselbaum2007}
\bibinfo{author}{\bibfnamefont{A.}~\bibnamefont{Weichselbaum}}
  \bibnamefont{and} \bibinfo{author}{\bibfnamefont{J.}~\bibnamefont{von
  Delft}}, \bibinfo{journal}{Phys. Rev. Lett.} \textbf{\bibinfo{volume}{99}},
  \bibinfo{pages}{076402} (\bibinfo{year}{2007}).

\bibitem[{\citenamefont{\v{Z}itko}(2011{\natexlab{a}})}]{zitko2011}
\bibinfo{author}{\bibfnamefont{R.}~\bibnamefont{\v{Z}itko}},
  \bibinfo{journal}{Phys. Rev. B} \textbf{\bibinfo{volume}{84}},
  \bibinfo{pages}{085142} (\bibinfo{year}{2011}{\natexlab{a}}).

\bibitem[{\citenamefont{Freyn and Florens}(2009)}]{freyn2009}
\bibinfo{author}{\bibfnamefont{A.}~\bibnamefont{Freyn}} \bibnamefont{and}
  \bibinfo{author}{\bibfnamefont{S.}~\bibnamefont{Florens}},
  \textbf{\bibinfo{volume}{79}}, \bibinfo{pages}{121102}
  (\bibinfo{year}{2009}).

\bibitem[{\citenamefont{Bulla et~al.}(2001)\citenamefont{Bulla, Costi, and
  Vollhardt}}]{bulla2001}
\bibinfo{author}{\bibfnamefont{R.}~\bibnamefont{Bulla}},
  \bibinfo{author}{\bibfnamefont{T.~A.} \bibnamefont{Costi}}, \bibnamefont{and}
  \bibinfo{author}{\bibfnamefont{D.}~\bibnamefont{Vollhardt}},
  \bibinfo{journal}{Phys. Rev. B} \textbf{\bibinfo{volume}{64}},
  \bibinfo{pages}{045103} (\bibinfo{year}{2001}).

\bibitem[{\citenamefont{Coleman}(2002)}]{coleman2002}
\bibinfo{author}{\bibfnamefont{P.}~\bibnamefont{Coleman}},
  \emph{\bibinfo{title}{Local moment physics in heavy electron systems}},
  \bibinfo{howpublished}{cond-mat/0206003} (\bibinfo{year}{2002}).

\bibitem[{\citenamefont{Allen}(2005)}]{allen2005}
\bibinfo{author}{\bibfnamefont{J.~W.} \bibnamefont{Allen}},
  \bibinfo{journal}{J. Phys. Soc. Jpn.} \textbf{\bibinfo{volume}{74}},
  \bibinfo{pages}{34} (\bibinfo{year}{2005}).

\bibitem[{\citenamefont{Glazman and Raikh}(1988)}]{glazman1988}
\bibinfo{author}{\bibfnamefont{L.~I.} \bibnamefont{Glazman}} \bibnamefont{and}
  \bibinfo{author}{\bibfnamefont{M.~E.} \bibnamefont{Raikh}},
  \bibinfo{journal}{JETP Lett.} \textbf{\bibinfo{volume}{47}},
  \bibinfo{pages}{452} (\bibinfo{year}{1988}).

\bibitem[{\citenamefont{Goldhaber-Gordon
  et~al.}(1998{\natexlab{a}})\citenamefont{Goldhaber-Gordon, Shtrikman, Mahalu,
  Abusch-Magder, Meirav, and Kastner}}]{goldhabergordon1998b}
\bibinfo{author}{\bibfnamefont{D.}~\bibnamefont{Goldhaber-Gordon}},
  \bibinfo{author}{\bibfnamefont{H.}~\bibnamefont{Shtrikman}},
  \bibinfo{author}{\bibfnamefont{D.}~\bibnamefont{Mahalu}},
  \bibinfo{author}{\bibfnamefont{D.}~\bibnamefont{Abusch-Magder}},
  \bibinfo{author}{\bibfnamefont{U.}~\bibnamefont{Meirav}}, \bibnamefont{and}
  \bibinfo{author}{\bibfnamefont{M.~A.} \bibnamefont{Kastner}},
  \bibinfo{journal}{Nature} \textbf{\bibinfo{volume}{391}},
  \bibinfo{pages}{156} (\bibinfo{year}{1998}{\natexlab{a}}).

\bibitem[{\citenamefont{Goldhaber-Gordon
  et~al.}(1998{\natexlab{b}})\citenamefont{Goldhaber-Gordon, G\"ores, Kastner,
  Shtrikman, Mahalu, and Meirav}}]{goldhabergordon1998a}
\bibinfo{author}{\bibfnamefont{D.}~\bibnamefont{Goldhaber-Gordon}},
  \bibinfo{author}{\bibfnamefont{J.}~\bibnamefont{G\"ores}},
  \bibinfo{author}{\bibfnamefont{M.~A.} \bibnamefont{Kastner}},
  \bibinfo{author}{\bibfnamefont{H.}~\bibnamefont{Shtrikman}},
  \bibinfo{author}{\bibfnamefont{D.}~\bibnamefont{Mahalu}}, \bibnamefont{and}
  \bibinfo{author}{\bibfnamefont{U.}~\bibnamefont{Meirav}},
  \bibinfo{journal}{Phys. Rev. Lett.} \textbf{\bibinfo{volume}{81}},
  \bibinfo{pages}{5225} (\bibinfo{year}{1998}{\natexlab{b}}).

\bibitem[{\citenamefont{Cronenwett et~al.}(1998)\citenamefont{Cronenwett,
  Oosterkamp, and Kouwenhoven}}]{cronenwett1998}
\bibinfo{author}{\bibfnamefont{S.~M.} \bibnamefont{Cronenwett}},
  \bibinfo{author}{\bibfnamefont{T.~H.} \bibnamefont{Oosterkamp}},
  \bibnamefont{and} \bibinfo{author}{\bibfnamefont{L.~P.}
  \bibnamefont{Kouwenhoven}}, \bibinfo{journal}{Science}
  \textbf{\bibinfo{volume}{281}}, \bibinfo{pages}{540} (\bibinfo{year}{1998}).

\bibitem[{\citenamefont{van~der Wiel et~al.}(2000)\citenamefont{van~der Wiel,
  Franceschi, Fujisawa, Elzerman, Tarucha, and Kouwenhoven}}]{wiel2000}
\bibinfo{author}{\bibfnamefont{W.~G.} \bibnamefont{van~der Wiel}},
  \bibinfo{author}{\bibfnamefont{S.~D.} \bibnamefont{Franceschi}},
  \bibinfo{author}{\bibfnamefont{T.}~\bibnamefont{Fujisawa}},
  \bibinfo{author}{\bibfnamefont{J.~M.} \bibnamefont{Elzerman}},
  \bibinfo{author}{\bibfnamefont{S.}~\bibnamefont{Tarucha}}, \bibnamefont{and}
  \bibinfo{author}{\bibfnamefont{L.~P.} \bibnamefont{Kouwenhoven}},
  \bibinfo{journal}{Science} \textbf{\bibinfo{volume}{289}},
  \bibinfo{pages}{2105} (\bibinfo{year}{2000}).

\bibitem[{\citenamefont{Pustilnik and Glazman}(2004)}]{pustilnik2004}
\bibinfo{author}{\bibfnamefont{M.}~\bibnamefont{Pustilnik}} \bibnamefont{and}
  \bibinfo{author}{\bibfnamefont{L.}~\bibnamefont{Glazman}},
  \bibinfo{journal}{J. Phys.: Condens. Matter} \textbf{\bibinfo{volume}{16}},
  \bibinfo{pages}{R513} (\bibinfo{year}{2004}).

\bibitem[{\citenamefont{Potok et~al.}(2007)\citenamefont{Potok, Rau, Shtrikman,
  Oreg, and Goldhaber-Gordon}}]{potok2007}
\bibinfo{author}{\bibfnamefont{R.~M.} \bibnamefont{Potok}},
  \bibinfo{author}{\bibfnamefont{I.~G.} \bibnamefont{Rau}},
  \bibinfo{author}{\bibfnamefont{H.}~\bibnamefont{Shtrikman}},
  \bibinfo{author}{\bibfnamefont{Y.}~\bibnamefont{Oreg}}, \bibnamefont{and}
  \bibinfo{author}{\bibfnamefont{D.}~\bibnamefont{Goldhaber-Gordon}},
  \bibinfo{journal}{Nature} \textbf{\bibinfo{volume}{446}},
  \bibinfo{pages}{167} (\bibinfo{year}{2007}).

\bibitem[{\citenamefont{Andergassen et~al.}(2010)\citenamefont{Andergassen,
  Meden, Schoeller, Splettstoesser, and Wegewijs}}]{andergassen2010}
\bibinfo{author}{\bibfnamefont{S.}~\bibnamefont{Andergassen}},
  \bibinfo{author}{\bibfnamefont{V.}~\bibnamefont{Meden}},
  \bibinfo{author}{\bibfnamefont{H.}~\bibnamefont{Schoeller}},
  \bibinfo{author}{\bibfnamefont{J.}~\bibnamefont{Splettstoesser}},
  \bibnamefont{and} \bibinfo{author}{\bibfnamefont{M.~R.}
  \bibnamefont{Wegewijs}}, \bibinfo{journal}{Nanotechnology}
  \textbf{\bibinfo{volume}{21}}, \bibinfo{pages}{272001}
  (\bibinfo{year}{2010}).

\bibitem[{\citenamefont{Nygard et~al.}(2000)\citenamefont{Nygard, Cobden, and
  Lindelof}}]{nygard2000}
\bibinfo{author}{\bibfnamefont{J.}~\bibnamefont{Nygard}},
  \bibinfo{author}{\bibfnamefont{D.~H.} \bibnamefont{Cobden}},
  \bibnamefont{and} \bibinfo{author}{\bibfnamefont{P.~E.}
  \bibnamefont{Lindelof}}, \bibinfo{journal}{Nature}
  \textbf{\bibinfo{volume}{408}}, \bibinfo{pages}{342} (\bibinfo{year}{2000}).

\bibitem[{\citenamefont{Park et~al.}(2002)\citenamefont{Park, Pasupathy,
  Goldsmith, Chang, Yaish, Petta, Rinkoski, Sethna, Abruna, McEuen
  et~al.}}]{park2002}
\bibinfo{author}{\bibfnamefont{J.}~\bibnamefont{Park}},
  \bibinfo{author}{\bibfnamefont{A.~N.} \bibnamefont{Pasupathy}},
  \bibinfo{author}{\bibfnamefont{J.~I.} \bibnamefont{Goldsmith}},
  \bibinfo{author}{\bibfnamefont{C.}~\bibnamefont{Chang}},
  \bibinfo{author}{\bibfnamefont{Y.}~\bibnamefont{Yaish}},
  \bibinfo{author}{\bibfnamefont{J.~R.} \bibnamefont{Petta}},
  \bibinfo{author}{\bibfnamefont{M.}~\bibnamefont{Rinkoski}},
  \bibinfo{author}{\bibfnamefont{J.~P.} \bibnamefont{Sethna}},
  \bibinfo{author}{\bibfnamefont{H.~D.} \bibnamefont{Abruna}},
  \bibinfo{author}{\bibfnamefont{P.~L.} \bibnamefont{McEuen}},
  \bibnamefont{et~al.}, \bibinfo{journal}{Nature}
  \textbf{\bibinfo{volume}{417}}, \bibinfo{pages}{722} (\bibinfo{year}{2002}).

\bibitem[{\citenamefont{Liang et~al.}(2002)\citenamefont{Liang, Shores,
  Bockrath, Long, and Park}}]{liang2002}
\bibinfo{author}{\bibfnamefont{W.}~\bibnamefont{Liang}},
  \bibinfo{author}{\bibfnamefont{M.~P.} \bibnamefont{Shores}},
  \bibinfo{author}{\bibfnamefont{M.}~\bibnamefont{Bockrath}},
  \bibinfo{author}{\bibfnamefont{J.~R.} \bibnamefont{Long}}, \bibnamefont{and}
  \bibinfo{author}{\bibfnamefont{K.}~\bibnamefont{Park}},
  \bibinfo{journal}{Nature} \textbf{\bibinfo{volume}{417}},
  \bibinfo{pages}{725} (\bibinfo{year}{2002}).

\bibitem[{\citenamefont{Parks et~al.}(2010)\citenamefont{Parks, Champagne,
  Costi, Shum, Pasupathy, Neuscamman, Flores-Torres, Cornaglia, Aligia,
  Balseiro et~al.}}]{parks2010}
\bibinfo{author}{\bibfnamefont{J.~J.} \bibnamefont{Parks}},
  \bibinfo{author}{\bibfnamefont{A.~R.} \bibnamefont{Champagne}},
  \bibinfo{author}{\bibfnamefont{T.~A.} \bibnamefont{Costi}},
  \bibinfo{author}{\bibfnamefont{W.~W.} \bibnamefont{Shum}},
  \bibinfo{author}{\bibfnamefont{A.~N.} \bibnamefont{Pasupathy}},
  \bibinfo{author}{\bibfnamefont{E.}~\bibnamefont{Neuscamman}},
  \bibinfo{author}{\bibfnamefont{S.}~\bibnamefont{Flores-Torres}},
  \bibinfo{author}{\bibfnamefont{P.~S.} \bibnamefont{Cornaglia}},
  \bibinfo{author}{\bibfnamefont{A.~A.} \bibnamefont{Aligia}},
  \bibinfo{author}{\bibfnamefont{C.~A.} \bibnamefont{Balseiro}},
  \bibnamefont{et~al.}, \bibinfo{journal}{Science}
  \textbf{\bibinfo{volume}{328}}, \bibinfo{pages}{1370} (\bibinfo{year}{2010}).

\bibitem[{\citenamefont{Scott and Natelson}(2010)}]{scott2010}
\bibinfo{author}{\bibfnamefont{G.~D.} \bibnamefont{Scott}} \bibnamefont{and}
  \bibinfo{author}{\bibfnamefont{D.}~\bibnamefont{Natelson}},
  \bibinfo{journal}{ACS Nano} \textbf{\bibinfo{volume}{4}},
  \bibinfo{pages}{3560} (\bibinfo{year}{2010}).

\bibitem[{\citenamefont{Romeike
  et~al.}(2006{\natexlab{a}})\citenamefont{Romeike, Wegewijs, Hofstetter, and
  Schoeller}}]{romeike2006}
\bibinfo{author}{\bibfnamefont{C.}~\bibnamefont{Romeike}},
  \bibinfo{author}{\bibfnamefont{M.~R.} \bibnamefont{Wegewijs}},
  \bibinfo{author}{\bibfnamefont{W.}~\bibnamefont{Hofstetter}},
  \bibnamefont{and}
  \bibinfo{author}{\bibfnamefont{H.}~\bibnamefont{Schoeller}},
  \bibinfo{journal}{Phys. Rev. Lett.} \textbf{\bibinfo{volume}{96}},
  \bibinfo{pages}{196601} (\bibinfo{year}{2006}{\natexlab{a}}).

\bibitem[{\citenamefont{Romeike
  et~al.}(2006{\natexlab{b}})\citenamefont{Romeike, Wegewijs, Hofstetter, and
  Schoeller}}]{romeike2006b}
\bibinfo{author}{\bibfnamefont{C.}~\bibnamefont{Romeike}},
  \bibinfo{author}{\bibfnamefont{M.~R.} \bibnamefont{Wegewijs}},
  \bibinfo{author}{\bibfnamefont{W.}~\bibnamefont{Hofstetter}},
  \bibnamefont{and}
  \bibinfo{author}{\bibfnamefont{H.}~\bibnamefont{Schoeller}},
  \bibinfo{journal}{Phys. Rev. Lett.} \textbf{\bibinfo{volume}{97}},
  \bibinfo{pages}{206601} (\bibinfo{year}{2006}{\natexlab{b}}).

\bibitem[{\citenamefont{Madhavan et~al.}(1998)\citenamefont{Madhavan, Chen,
  Jamneala, Crommie, and Wingreen}}]{madhavan1998}
\bibinfo{author}{\bibfnamefont{V.}~\bibnamefont{Madhavan}},
  \bibinfo{author}{\bibfnamefont{W.}~\bibnamefont{Chen}},
  \bibinfo{author}{\bibfnamefont{T.}~\bibnamefont{Jamneala}},
  \bibinfo{author}{\bibfnamefont{M.}~\bibnamefont{Crommie}}, \bibnamefont{and}
  \bibinfo{author}{\bibfnamefont{N.~S.} \bibnamefont{Wingreen}},
  \bibinfo{journal}{Science} \textbf{\bibinfo{volume}{280}},
  \bibinfo{pages}{567} (\bibinfo{year}{1998}).

\bibitem[{\citenamefont{N\'eel et~al.}(2007)\citenamefont{N\'eel, Kr\"oger,
  Limot, Palotas, Hofer, and Berndt}}]{neel2007}
\bibinfo{author}{\bibfnamefont{N.}~\bibnamefont{N\'eel}},
  \bibinfo{author}{\bibfnamefont{J.}~\bibnamefont{Kr\"oger}},
  \bibinfo{author}{\bibfnamefont{L.}~\bibnamefont{Limot}},
  \bibinfo{author}{\bibfnamefont{K.}~\bibnamefont{Palotas}},
  \bibinfo{author}{\bibfnamefont{W.~A.} \bibnamefont{Hofer}}, \bibnamefont{and}
  \bibinfo{author}{\bibfnamefont{R.}~\bibnamefont{Berndt}},
  \bibinfo{journal}{Phys. Rev. Lett.} \textbf{\bibinfo{volume}{98}},
  \bibinfo{pages}{016801} (\bibinfo{year}{2007}).

\bibitem[{\citenamefont{Pr\"user et~al.}(2011)\citenamefont{Pr\"user,
  Wenderoth, Dargel, Weismann, Peters, Pruschke, and Ulbrich}}]{pruser2011}
\bibinfo{author}{\bibfnamefont{H.}~\bibnamefont{Pr\"user}},
  \bibinfo{author}{\bibfnamefont{M.}~\bibnamefont{Wenderoth}},
  \bibinfo{author}{\bibfnamefont{P.~E.} \bibnamefont{Dargel}},
  \bibinfo{author}{\bibfnamefont{A.}~\bibnamefont{Weismann}},
  \bibinfo{author}{\bibfnamefont{R.}~\bibnamefont{Peters}},
  \bibinfo{author}{\bibfnamefont{T.}~\bibnamefont{Pruschke}}, \bibnamefont{and}
  \bibinfo{author}{\bibfnamefont{R.~G.} \bibnamefont{Ulbrich}},
  \bibinfo{journal}{Nat. Phys.} \textbf{\bibinfo{volume}{7}},
  \bibinfo{pages}{203} (\bibinfo{year}{2011}).

\bibitem[{\citenamefont{Leggett et~al.}(1987)\citenamefont{Leggett,
  Chakravarty, Dorsey, Fisher, Garg, and Zwerger}}]{leggett1987}
\bibinfo{author}{\bibfnamefont{A.~J.} \bibnamefont{Leggett}},
  \bibinfo{author}{\bibfnamefont{S.}~\bibnamefont{Chakravarty}},
  \bibinfo{author}{\bibfnamefont{A.~T.} \bibnamefont{Dorsey}},
  \bibinfo{author}{\bibfnamefont{M.~P.~A.} \bibnamefont{Fisher}},
  \bibinfo{author}{\bibfnamefont{A.}~\bibnamefont{Garg}}, \bibnamefont{and}
  \bibinfo{author}{\bibfnamefont{W.}~\bibnamefont{Zwerger}},
  \bibinfo{journal}{Rev. Mod. Phys.} \textbf{\bibinfo{volume}{59}},
  \bibinfo{pages}{1} (\bibinfo{year}{1987}).

\bibitem[{\citenamefont{Metzner and Vollhardt}(1989)}]{metzner1989}
\bibinfo{author}{\bibfnamefont{W.}~\bibnamefont{Metzner}} \bibnamefont{and}
  \bibinfo{author}{\bibfnamefont{D.}~\bibnamefont{Vollhardt}},
  \bibinfo{journal}{Phys. Rev. Lett.} \textbf{\bibinfo{volume}{62}},
  \bibinfo{pages}{324} (\bibinfo{year}{1989}).

\bibitem[{\citenamefont{M\"uller-Hartmann}(1989)}]{mullerhartmann1989}
\bibinfo{author}{\bibfnamefont{E.}~\bibnamefont{M\"uller-Hartmann}},
  \bibinfo{journal}{Z. Phys. B: Condens. Matter} \textbf{\bibinfo{volume}{74}},
  \bibinfo{pages}{507} (\bibinfo{year}{1989}).

\bibitem[{\citenamefont{Kotliar}(2005)}]{kotliar2005}
\bibinfo{author}{\bibfnamefont{G.}~\bibnamefont{Kotliar}}, \bibinfo{journal}{J.
  Phys. Soc. Japan} \textbf{\bibinfo{volume}{74}}, \bibinfo{pages}{147}
  (\bibinfo{year}{2005}).

\bibitem[{\citenamefont{Maier et~al.}(2005)\citenamefont{Maier, Jarrell,
  Pruschke, and Hettler}}]{maier2005}
\bibinfo{author}{\bibfnamefont{T.}~\bibnamefont{Maier}},
  \bibinfo{author}{\bibfnamefont{M.}~\bibnamefont{Jarrell}},
  \bibinfo{author}{\bibfnamefont{T.}~\bibnamefont{Pruschke}}, \bibnamefont{and}
  \bibinfo{author}{\bibfnamefont{M.~H.} \bibnamefont{Hettler}},
  \bibinfo{journal}{Rev. Mod. Phys.} \textbf{\bibinfo{volume}{77}},
  \bibinfo{pages}{1027} (\bibinfo{year}{2005}).

\bibitem[{\citenamefont{Kotliar et~al.}(2006)\citenamefont{Kotliar, Savrasov,
  Haule, Oudovenko, Parcollet, and Marianetti}}]{kotliar2006}
\bibinfo{author}{\bibfnamefont{G.}~\bibnamefont{Kotliar}},
  \bibinfo{author}{\bibfnamefont{S.~Y.} \bibnamefont{Savrasov}},
  \bibinfo{author}{\bibfnamefont{K.}~\bibnamefont{Haule}},
  \bibinfo{author}{\bibfnamefont{V.~S.} \bibnamefont{Oudovenko}},
  \bibinfo{author}{\bibfnamefont{O.}~\bibnamefont{Parcollet}},
  \bibnamefont{and} \bibinfo{author}{\bibfnamefont{C.~A.}
  \bibnamefont{Marianetti}}, \bibinfo{journal}{Rev. Mod. Phys.}
  \textbf{\bibinfo{volume}{78}}, \bibinfo{pages}{865} (\bibinfo{year}{2006}).

\bibitem[{\citenamefont{Georges}(2011)}]{Georges:2011hr}
\bibinfo{author}{\bibfnamefont{A.}~\bibnamefont{Georges}},
  \bibinfo{journal}{Annalen der Physik} \textbf{\bibinfo{volume}{523}},
  \bibinfo{pages}{672} (\bibinfo{year}{2011}).

\bibitem[{\citenamefont{Werner et~al.}(2006)\citenamefont{Werner, Comanac, de'
  Medici, Troyer, and Millis}}]{werner2006}
\bibinfo{author}{\bibfnamefont{P.}~\bibnamefont{Werner}},
  \bibinfo{author}{\bibfnamefont{A.}~\bibnamefont{Comanac}},
  \bibinfo{author}{\bibfnamefont{L.}~\bibnamefont{de' Medici}},
  \bibinfo{author}{\bibfnamefont{M.}~\bibnamefont{Troyer}}, \bibnamefont{and}
  \bibinfo{author}{\bibfnamefont{A.~J.} \bibnamefont{Millis}},
  \bibinfo{journal}{Phys. Rev. Lett.} \textbf{\bibinfo{volume}{97}},
  \bibinfo{pages}{076405} (\bibinfo{year}{2006}).

\bibitem[{\citenamefont{Haule}(2007)}]{haule2007}
\bibinfo{author}{\bibfnamefont{K.}~\bibnamefont{Haule}},
  \bibinfo{journal}{Phys. Rev. B} \textbf{\bibinfo{volume}{75}},
  \bibinfo{pages}{155113} (\bibinfo{year}{2007}).

\bibitem[{\citenamefont{Gull et~al.}(2011)\citenamefont{Gull, Millis,
  Lichtenstein, Rubtsov, Troyer, and Werner}}]{gull2011}
\bibinfo{author}{\bibfnamefont{E.}~\bibnamefont{Gull}},
  \bibinfo{author}{\bibfnamefont{A.~J.} \bibnamefont{Millis}},
  \bibinfo{author}{\bibfnamefont{A.~I.} \bibnamefont{Lichtenstein}},
  \bibinfo{author}{\bibfnamefont{A.~N.} \bibnamefont{Rubtsov}},
  \bibinfo{author}{\bibfnamefont{M.}~\bibnamefont{Troyer}}, \bibnamefont{and}
  \bibinfo{author}{\bibfnamefont{P.}~\bibnamefont{Werner}},
  \bibinfo{journal}{Rev. Mod. Phys.} \textbf{\bibinfo{volume}{83}},
  \bibinfo{pages}{349} (\bibinfo{year}{2011}).

\bibitem[{\citenamefont{Jarrell and Gubernatis}(1996)}]{jarrell1996}
\bibinfo{author}{\bibfnamefont{M.}~\bibnamefont{Jarrell}} \bibnamefont{and}
  \bibinfo{author}{\bibfnamefont{J.~E.} \bibnamefont{Gubernatis}},
  \bibinfo{journal}{Physics Reports} \textbf{\bibinfo{volume}{269}},
  \bibinfo{pages}{133} (\bibinfo{year}{1996}).

\bibitem[{\citenamefont{Vidberg and Serene}(1977)}]{Vidberg:1977vo}
\bibinfo{author}{\bibfnamefont{H.~J.} \bibnamefont{Vidberg}} \bibnamefont{and}
  \bibinfo{author}{\bibfnamefont{J.~W.} \bibnamefont{Serene}},
  \bibinfo{journal}{Journal of Low Temperature Physics}
  \textbf{\bibinfo{volume}{29}}, \bibinfo{pages}{179} (\bibinfo{year}{1977}).

\bibitem[{\citenamefont{Beach et~al.}(2000)\citenamefont{Beach, Gooding, and
  Marsiglio}}]{PhysRevB.61.5147}
\bibinfo{author}{\bibfnamefont{K.~S.~D.} \bibnamefont{Beach}},
  \bibinfo{author}{\bibfnamefont{R.~J.} \bibnamefont{Gooding}},
  \bibnamefont{and}
  \bibinfo{author}{\bibfnamefont{F.}~\bibnamefont{Marsiglio}},
  \bibinfo{journal}{Phys. Rev. B} \textbf{\bibinfo{volume}{61}},
  \bibinfo{pages}{5147} (\bibinfo{year}{2000}),
  \urlprefix\url{http://link.aps.org/doi/10.1103/PhysRevB.61.5147}.

\bibitem[{\citenamefont{Jarrell and Pruschke}(1994)}]{Jarrell:1994ut}
\bibinfo{author}{\bibfnamefont{M.}~\bibnamefont{Jarrell}} \bibnamefont{and}
  \bibinfo{author}{\bibfnamefont{T.}~\bibnamefont{Pruschke}},
  \bibinfo{journal}{Physical Review B} \textbf{\bibinfo{volume}{49}},
  \bibinfo{pages}{1458} (\bibinfo{year}{1994}).

\bibitem[{\citenamefont{Jarrell et~al.}(1995)\citenamefont{Jarrell, Freericks,
  and Pruschke}}]{Jarrell:1995te}
\bibinfo{author}{\bibfnamefont{M.}~\bibnamefont{Jarrell}},
  \bibinfo{author}{\bibfnamefont{J.~K.} \bibnamefont{Freericks}},
  \bibnamefont{and} \bibinfo{author}{\bibfnamefont{T.}~\bibnamefont{Pruschke}},
  \bibinfo{journal}{Physical Review B} \textbf{\bibinfo{volume}{51}},
  \bibinfo{pages}{11704} (\bibinfo{year}{1995}).

\bibitem[{\citenamefont{Grenzenbach et~al.}(2006)\citenamefont{Grenzenbach,
  Anders, Czycholl, and Pruschke}}]{grenzenbach2006}
\bibinfo{author}{\bibfnamefont{C.}~\bibnamefont{Grenzenbach}},
  \bibinfo{author}{\bibfnamefont{F.~B.} \bibnamefont{Anders}},
  \bibinfo{author}{\bibfnamefont{G.}~\bibnamefont{Czycholl}}, \bibnamefont{and}
  \bibinfo{author}{\bibfnamefont{T.}~\bibnamefont{Pruschke}},
  \bibinfo{journal}{Phys. Rev. B} \textbf{\bibinfo{volume}{74}},
  \bibinfo{pages}{195119} (\bibinfo{year}{2006}).

\bibitem[{\citenamefont{Georges and Kotliar}(1992)}]{georges1992}
\bibinfo{author}{\bibfnamefont{A.}~\bibnamefont{Georges}} \bibnamefont{and}
  \bibinfo{author}{\bibfnamefont{G.}~\bibnamefont{Kotliar}},
  \bibinfo{journal}{Phys. Rev. B} \textbf{\bibinfo{volume}{45}},
  \bibinfo{pages}{6479} (\bibinfo{year}{1992}).

\bibitem[{\citenamefont{Georges and Krauth}(1992)}]{georges1992mott}
\bibinfo{author}{\bibfnamefont{A.}~\bibnamefont{Georges}} \bibnamefont{and}
  \bibinfo{author}{\bibfnamefont{W.}~\bibnamefont{Krauth}},
  \bibinfo{journal}{Phys. Rev. Lett.} \textbf{\bibinfo{volume}{69}},
  \bibinfo{pages}{1240} (\bibinfo{year}{1992}).

\bibitem[{\citenamefont{Rozenberg et~al.}(1992)\citenamefont{Rozenberg, Zhang,
  and Kotliar}}]{rozenberg1992}
\bibinfo{author}{\bibfnamefont{M.~J.} \bibnamefont{Rozenberg}},
  \bibinfo{author}{\bibfnamefont{X.~Y.} \bibnamefont{Zhang}}, \bibnamefont{and}
  \bibinfo{author}{\bibfnamefont{G.}~\bibnamefont{Kotliar}},
  \bibinfo{journal}{Phys. Rev. Lett.} \textbf{\bibinfo{volume}{69}},
  \bibinfo{pages}{1236} (\bibinfo{year}{1992}).

\bibitem[{\citenamefont{Jarrell}(1992)}]{jarrell1992dmft}
\bibinfo{author}{\bibfnamefont{M.}~\bibnamefont{Jarrell}},
  \bibinfo{journal}{Phys. Rev. Lett.} \textbf{\bibinfo{volume}{69}},
  \bibinfo{pages}{168} (\bibinfo{year}{1992}).

\bibitem[{\citenamefont{Hubbard}(1963)}]{hubbard1963}
\bibinfo{author}{\bibfnamefont{J.}~\bibnamefont{Hubbard}},
  \bibinfo{journal}{Proc. R. Soc. London, Ser. A}
  \textbf{\bibinfo{volume}{276}}, \bibinfo{pages}{238} (\bibinfo{year}{1963}).

\bibitem[{\citenamefont{Kanamori}(1963)}]{kanamori1963}
\bibinfo{author}{\bibfnamefont{J.}~\bibnamefont{Kanamori}},
  \bibinfo{journal}{Prog. Theor. Phys.} \textbf{\bibinfo{volume}{30}},
  \bibinfo{pages}{275} (\bibinfo{year}{1963}).

\bibitem[{\citenamefont{Gutzwiller}(1963)}]{gutzwiller1963}
\bibinfo{author}{\bibfnamefont{M.~C.} \bibnamefont{Gutzwiller}},
  \bibinfo{journal}{Phys. Rev. Lett.} \textbf{\bibinfo{volume}{10}},
  \bibinfo{pages}{159} (\bibinfo{year}{1963}).

\bibitem[{\citenamefont{Anderson}(1961)}]{anderson1961}
\bibinfo{author}{\bibfnamefont{P.~W.} \bibnamefont{Anderson}},
  \bibinfo{journal}{Phys. Rev.} \textbf{\bibinfo{volume}{124}},
  \bibinfo{pages}{41} (\bibinfo{year}{1961}).

\bibitem[{\citenamefont{Anderson}(1967)}]{anderson1967}
\bibinfo{author}{\bibfnamefont{P.~W.} \bibnamefont{Anderson}},
  \bibinfo{journal}{Phys. Rev.} \textbf{\bibinfo{volume}{164}},
  \bibinfo{pages}{352} (\bibinfo{year}{1967}).

\bibitem[{\citenamefont{Anderson}(1978)}]{anderson1978}
\bibinfo{author}{\bibfnamefont{P.~W.} \bibnamefont{Anderson}},
  \bibinfo{journal}{Rev. Mod. Phys.} \textbf{\bibinfo{volume}{50}},
  \bibinfo{pages}{191} (\bibinfo{year}{1978}).

\bibitem[{\citenamefont{Sakai and Kuramoto}(1994)}]{sakai1994}
\bibinfo{author}{\bibfnamefont{O.}~\bibnamefont{Sakai}} \bibnamefont{and}
  \bibinfo{author}{\bibfnamefont{Y.}~\bibnamefont{Kuramoto}},
  \bibinfo{journal}{Solid State Commun.} \textbf{\bibinfo{volume}{89}},
  \bibinfo{pages}{307} (\bibinfo{year}{1994}).

\bibitem[{\citenamefont{Bulla}(1999)}]{bulla1999}
\bibinfo{author}{\bibfnamefont{R.}~\bibnamefont{Bulla}},
  \bibinfo{journal}{Phys. Rev. Lett.} \textbf{\bibinfo{volume}{83}},
  \bibinfo{pages}{136} (\bibinfo{year}{1999}).

\bibitem[{\citenamefont{Bulla et~al.}(1997)\citenamefont{Bulla, Pruschke, and
  Hewson}}]{bulla1997}
\bibinfo{author}{\bibfnamefont{R.}~\bibnamefont{Bulla}},
  \bibinfo{author}{\bibfnamefont{T.}~\bibnamefont{Pruschke}}, \bibnamefont{and}
  \bibinfo{author}{\bibfnamefont{A.~C.} \bibnamefont{Hewson}},
  \bibinfo{journal}{J. Phys.: Condens. Matter.} \textbf{\bibinfo{volume}{9}},
  \bibinfo{pages}{10463} (\bibinfo{year}{1997}).

\bibitem[{\citenamefont{Georges et~al.}(1996)\citenamefont{Georges, Kotliar,
  Krauth, and Rozenberg}}]{georges1996}
\bibinfo{author}{\bibfnamefont{A.}~\bibnamefont{Georges}},
  \bibinfo{author}{\bibfnamefont{G.}~\bibnamefont{Kotliar}},
  \bibinfo{author}{\bibfnamefont{W.}~\bibnamefont{Krauth}}, \bibnamefont{and}
  \bibinfo{author}{\bibfnamefont{M.~J.} \bibnamefont{Rozenberg}},
  \bibinfo{journal}{Rev. Mod. Phys.} \textbf{\bibinfo{volume}{68}},
  \bibinfo{pages}{13} (\bibinfo{year}{1996}).

\bibitem[{\citenamefont{\v{Z}itko}(2009{\natexlab{a}})}]{broyden}
\bibinfo{author}{\bibfnamefont{R.}~\bibnamefont{\v{Z}itko}},
  \bibinfo{journal}{Phys. Rev. B} \textbf{\bibinfo{volume}{80}},
  \bibinfo{pages}{125125} (\bibinfo{year}{2009}{\natexlab{a}}).

\bibitem[{\citenamefont{\v{Z}itko}(2009{\natexlab{b}})}]{odesolv}
\bibinfo{author}{\bibfnamefont{R.}~\bibnamefont{\v{Z}itko}},
  \bibinfo{journal}{Comp. Phys. Comm.} \textbf{\bibinfo{volume}{180}},
  \bibinfo{pages}{1271} (\bibinfo{year}{2009}{\natexlab{b}}).

\bibitem[{\citenamefont{Campo and Oliveira}(2004)}]{campo2004}
\bibinfo{author}{\bibfnamefont{V.~L.} \bibnamefont{Campo}} \bibnamefont{and}
  \bibinfo{author}{\bibfnamefont{L.~N.} \bibnamefont{Oliveira}},
  \bibinfo{journal}{Phys. Rev. B} \textbf{\bibinfo{volume}{70}},
  \bibinfo{pages}{153401} (\bibinfo{year}{2004}).

\bibitem[{\citenamefont{\v{Z}itko}(2006)}]{nrglj}
\bibinfo{author}{\bibfnamefont{R.}~\bibnamefont{\v{Z}itko}},
  \emph{\bibinfo{title}{{NRG} {L}jubljana}},
  \bibinfo{howpublished}{\url{http://nrgljubljana.ijs.si/}}
  (\bibinfo{year}{2006}).

\bibitem[{\citenamefont{\v{Z}itko}(2011{\natexlab{b}})}]{sneg}
\bibinfo{author}{\bibfnamefont{R.}~\bibnamefont{\v{Z}itko}},
  \bibinfo{journal}{Comp. Phys. Comm.} \textbf{\bibinfo{volume}{1982}},
  \bibinfo{pages}{2259} (\bibinfo{year}{2011}{\natexlab{b}}).

\bibitem[{\citenamefont{\v{Z}itko}(2011{\natexlab{c}})}]{errors}
\bibinfo{author}{\bibfnamefont{R.}~\bibnamefont{\v{Z}itko}},
  \bibinfo{journal}{Phys. Rev. B} \textbf{\bibinfo{volume}{84}},
  \bibinfo{pages}{085142} (\bibinfo{year}{2011}{\natexlab{c}}).

\bibitem[{\citenamefont{\v{Z}itko and Bon\v{c}a}(2006)}]{sidecoupled}
\bibinfo{author}{\bibfnamefont{R.}~\bibnamefont{\v{Z}itko}} \bibnamefont{and}
  \bibinfo{author}{\bibfnamefont{J.}~\bibnamefont{Bon\v{c}a}},
  \bibinfo{journal}{Phys. Rev. B} \textbf{\bibinfo{volume}{73}},
  \bibinfo{pages}{035332} (\bibinfo{year}{2006}).

\bibitem[{\citenamefont{T\'oth et~al.}(2008)\citenamefont{T\'oth, Moca, Legeza,
  and Zar\'and}}]{toth2008nrg}
\bibinfo{author}{\bibfnamefont{A.~I.} \bibnamefont{T\'oth}},
  \bibinfo{author}{\bibfnamefont{C.~P.} \bibnamefont{Moca}},
  \bibinfo{author}{\bibfnamefont{O.}~\bibnamefont{Legeza}}, \bibnamefont{and}
  \bibinfo{author}{\bibfnamefont{G.}~\bibnamefont{Zar\'and}},
  \bibinfo{journal}{Phys. Rev. B} \textbf{\bibinfo{volume}{78}},
  \bibinfo{pages}{245109} (\bibinfo{year}{2008}).

\bibitem[{\citenamefont{Weichselbaum}(2012)}]{Weichselbaum:2012tq}
\bibinfo{author}{\bibfnamefont{A.}~\bibnamefont{Weichselbaum}},
  \bibinfo{journal}{Annals of Physics} \textbf{\bibinfo{volume}{327}},
  \bibinfo{pages}{2972} (\bibinfo{year}{2012}).

\bibitem[{\citenamefont{{Hafermann} et~al.}(2012)\citenamefont{{Hafermann},
  {Patton}, and {Werner}}}]{2012PhRvB..85t5106H}
\bibinfo{author}{\bibfnamefont{H.}~\bibnamefont{{Hafermann}}},
  \bibinfo{author}{\bibfnamefont{K.~R.} \bibnamefont{{Patton}}},
  \bibnamefont{and} \bibinfo{author}{\bibfnamefont{P.}~\bibnamefont{{Werner}}},
  \textbf{\bibinfo{volume}{85}}, \bibinfo{eid}{205106} (\bibinfo{year}{2012}),
  \eprint{1108.1936}.

\bibitem[{\citenamefont{Silver et~al.}(1990)\citenamefont{Silver, Sivia, and
  Gubernatis}}]{PhysRevB.41.2380}
\bibinfo{author}{\bibfnamefont{R.~N.} \bibnamefont{Silver}},
  \bibinfo{author}{\bibfnamefont{D.~S.} \bibnamefont{Sivia}}, \bibnamefont{and}
  \bibinfo{author}{\bibfnamefont{J.~E.} \bibnamefont{Gubernatis}},
  \bibinfo{journal}{Phys. Rev. B} \textbf{\bibinfo{volume}{41}},
  \bibinfo{pages}{2380} (\bibinfo{year}{1990}),
  \urlprefix\url{http://link.aps.org/doi/10.1103/PhysRevB.41.2380}.

\bibitem[{\citenamefont{Haldane}(1977)}]{haldane1977}
\bibinfo{author}{\bibfnamefont{F.~D.~M.} \bibnamefont{Haldane}},
  \bibinfo{journal}{Phys. Rev. B} \textbf{\bibinfo{volume}{15}},
  \bibinfo{pages}{281} (\bibinfo{year}{1977}).

\bibitem[{\citenamefont{Schrieffer and Wolff}(1966)}]{schrieffer1966}
\bibinfo{author}{\bibfnamefont{J.~R.} \bibnamefont{Schrieffer}}
  \bibnamefont{and} \bibinfo{author}{\bibfnamefont{P.~A.} \bibnamefont{Wolff}},
  \bibinfo{journal}{Phys. Rev.} \textbf{\bibinfo{volume}{149}},
  \bibinfo{pages}{491} (\bibinfo{year}{1966}).

\bibitem[{\citenamefont{Suhl}(1965)}]{suhl1965}
\bibinfo{author}{\bibfnamefont{H.}~\bibnamefont{Suhl}}, \bibinfo{journal}{Phys.
  Rev.} \textbf{\bibinfo{volume}{138}}, \bibinfo{pages}{A515}
  (\bibinfo{year}{1965}).

\bibitem[{\citenamefont{Langreth}(1966)}]{langreth1966}
\bibinfo{author}{\bibfnamefont{D.~C.} \bibnamefont{Langreth}},
  \bibinfo{journal}{Phys. Rev.} \textbf{\bibinfo{volume}{150}},
  \bibinfo{pages}{516} (\bibinfo{year}{1966}).

\bibitem[{\citenamefont{Frota}(1992)}]{frota1992}
\bibinfo{author}{\bibfnamefont{H.~O.} \bibnamefont{Frota}},
  \bibinfo{journal}{Phys. Rev. B} \textbf{\bibinfo{volume}{45}},
  \bibinfo{pages}{1096} (\bibinfo{year}{1992}).

\bibitem[{\citenamefont{Bulla et~al.}(2000)\citenamefont{Bulla, Glossop, Logan,
  and Pruschke}}]{bulla2000}
\bibinfo{author}{\bibfnamefont{R.}~\bibnamefont{Bulla}},
  \bibinfo{author}{\bibfnamefont{M.~T.} \bibnamefont{Glossop}},
  \bibinfo{author}{\bibfnamefont{D.~E.} \bibnamefont{Logan}}, \bibnamefont{and}
  \bibinfo{author}{\bibfnamefont{T.}~\bibnamefont{Pruschke}},
  \bibinfo{journal}{J. Phys.: Condens. Matter} \textbf{\bibinfo{volume}{12}},
  \bibinfo{pages}{4899} (\bibinfo{year}{2000}).

\bibitem[{\citenamefont{Logan et~al.}(1998)\citenamefont{Logan, Eastwood, and
  Tusch}}]{logan1998}
\bibinfo{author}{\bibfnamefont{D.~E.} \bibnamefont{Logan}},
  \bibinfo{author}{\bibfnamefont{M.~P.} \bibnamefont{Eastwood}},
  \bibnamefont{and} \bibinfo{author}{\bibfnamefont{M.~A.} \bibnamefont{Tusch}},
  \bibinfo{journal}{J. Phys.: Condens. Matter} \textbf{\bibinfo{volume}{10}},
  \bibinfo{pages}{2673} (\bibinfo{year}{1998}).

\bibitem[{\citenamefont{Dickens and Logan}(2001)}]{dickens2001}
\bibinfo{author}{\bibfnamefont{N.~L.} \bibnamefont{Dickens}} \bibnamefont{and}
  \bibinfo{author}{\bibfnamefont{D.~E.} \bibnamefont{Logan}},
  \bibinfo{journal}{J. Phys.: Condens. Matter} \textbf{\bibinfo{volume}{13}},
  \bibinfo{pages}{4505} (\bibinfo{year}{2001}).

\bibitem[{\citenamefont{Meir and Wingreen}(1992)}]{meir1992}
\bibinfo{author}{\bibfnamefont{Y.}~\bibnamefont{Meir}} \bibnamefont{and}
  \bibinfo{author}{\bibfnamefont{N.~S.} \bibnamefont{Wingreen}},
  \bibinfo{journal}{Phys. Rev. Lett.} \textbf{\bibinfo{volume}{68}},
  \bibinfo{pages}{2512} (\bibinfo{year}{1992}).

\bibitem[{\citenamefont{Hewson}(2006)}]{hewson2006spin}
\bibinfo{author}{\bibfnamefont{A.~C.} \bibnamefont{Hewson}},
  \bibinfo{journal}{J. Phys.: Condens. Matter} \textbf{\bibinfo{volume}{18}},
  \bibinfo{pages}{1815} (\bibinfo{year}{2006}).

\bibitem[{\citenamefont{Shiba}(1975)}]{shiba1975}
\bibinfo{author}{\bibfnamefont{H.}~\bibnamefont{Shiba}},
  \bibinfo{journal}{Prog. Theor. Phys.} \textbf{\bibinfo{volume}{54}},
  \bibinfo{pages}{967} (\bibinfo{year}{1975}).

\bibitem[{\citenamefont{Merker et~al.}(2012)\citenamefont{Merker, Weichselbaum,
  and Costi}}]{Merker:2012fb}
\bibinfo{author}{\bibfnamefont{L.}~\bibnamefont{Merker}},
  \bibinfo{author}{\bibfnamefont{A.}~\bibnamefont{Weichselbaum}},
  \bibnamefont{and} \bibinfo{author}{\bibfnamefont{T.~A.} \bibnamefont{Costi}},
  \bibinfo{journal}{Physical Review B} \textbf{\bibinfo{volume}{86}},
  \bibinfo{pages}{075153} (\bibinfo{year}{2012}).

\bibitem[{\citenamefont{Karski et~al.}(2005)\citenamefont{Karski, Raas, and
  Uhrig}}]{karski2005}
\bibinfo{author}{\bibfnamefont{M.}~\bibnamefont{Karski}},
  \bibinfo{author}{\bibfnamefont{C.}~\bibnamefont{Raas}}, \bibnamefont{and}
  \bibinfo{author}{\bibfnamefont{G.~S.} \bibnamefont{Uhrig}},
  \bibinfo{journal}{Phys. Rev. B} \textbf{\bibinfo{volume}{72}},
  \bibinfo{pages}{113110} (\bibinfo{year}{2005}).

\bibitem[{\citenamefont{Karski et~al.}(2008)\citenamefont{Karski, Raas, and
  Uhrig}}]{karski2008}
\bibinfo{author}{\bibfnamefont{M.}~\bibnamefont{Karski}},
  \bibinfo{author}{\bibfnamefont{C.}~\bibnamefont{Raas}}, \bibnamefont{and}
  \bibinfo{author}{\bibfnamefont{G.~S.} \bibnamefont{Uhrig}},
  \bibinfo{journal}{Phys. Rev. B} \textbf{\bibinfo{volume}{77}},
  \bibinfo{pages}{075116} (\bibinfo{year}{2008}).

\bibitem[{\citenamefont{Deng et~al.}(2012)\citenamefont{Deng, Mravlje, Zitko,
  Ferrero, Kotliar, and Georges}}]{deng}
\bibinfo{author}{\bibfnamefont{X.}~\bibnamefont{Deng}},
  \bibinfo{author}{\bibfnamefont{J.}~\bibnamefont{Mravlje}},
  \bibinfo{author}{\bibfnamefont{R.}~\bibnamefont{Zitko}},
  \bibinfo{author}{\bibfnamefont{M.}~\bibnamefont{Ferrero}},
  \bibinfo{author}{\bibfnamefont{G.}~\bibnamefont{Kotliar}}, \bibnamefont{and}
  \bibinfo{author}{\bibfnamefont{A.}~\bibnamefont{Georges}},
  \emph{\bibinfo{title}{How bad metals turn good: spectroscopic signatures of
  resilient quasiparticles}}, \bibinfo{howpublished}{cond-mat/1210.1769}
  (\bibinfo{year}{2012}).

\end{thebibliography}

\end{document}